\documentclass[pdflatex,sn-mathphys-num]{sn-jnl}

\usepackage{aas_macros}
\usepackage{graphicx}%
\usepackage{multirow}%
\usepackage{amsmath,amssymb,amsfonts}%
\usepackage{amsthm}%
\usepackage{mathrsfs}%
\usepackage[title]{appendix}%
\usepackage{xcolor}%
\usepackage{textcomp}%
\usepackage{manyfoot}%
\usepackage{booktabs}%
\usepackage{algorithm}%
\usepackage{algorithmicx}%
\usepackage{algpseudocode}%
\usepackage{listings}%
\usepackage[utf8]{inputenc}
\usepackage{hyperref}
\usepackage{natbib}
\usepackage{float}
\providecommand{\teff}{\ensuremath{T_{\rm eff}}}

\begin{document}

\title[The PLATO Science Calibration and Validation Plan]{The PLATO Science Calibration and Validation Plan: Targets for the First Long-pointing Field.}

\author*[1]{\fnm{Konstanze} \sur{Zwintz}}\email{konstanze.zwintz@uibk.ac.at}

\author[2]{\fnm{Conny} \sur{Aerts}}

\author[2]{\fnm{Andrew} \sur{Tkachenko}}

\author[3]{\fnm{Juan} \sur{Cabrera}}

\author[4]{\fnm{Orlagh} \sur{Creevey}}

\author[5]{\fnm{René} \sur{Heller}}

\author[2]{\fnm{Nicholas} \sur{Jannsen}}

\author[5]{\fnm{Chen} \sur{Jiang}}

\author[6]{\fnm{Oleg} \sur{Kochukhov}}

\author[7]{\fnm{Antonino Francesco}\sur{Lanza}}

\author[8]{\fnm{Pierre F. L.} \sur{Maxted}}

\author[7]{\fnm{Sergio} \sur{Messina}}

\author[9]{\fnm{Andrea} \sur{Miglio}}

\author[10]{\fnm{Thierry} \sur{Morel}}

\author[11]{\fnm{Benoît} \sur{Mosser}}

\author[12]{\fnm{Rhita} \sur{Ouazzani}}

\author[8]{\fnm{John} \sur{Southworth}}


\author[5]{\fnm{Matthias} \sur{Ammler van-Eiff}}

\author[13]{\fnm{Jeroen} \sur{Audenaert}}

\author[14,15]{\fnm{Paul G.} \sur{Beck}}

\author[12]{\fnm{Kevin} \sur{Belkacem}}

\author[5]{\fnm{Aaron} \sur{Birch}}

\author[16]{\fnm{Diego} \sur{Bossini}}

\author[17]{\fnm{Angela} \sur{Bragaglia}}

\author[9,17]{\fnm{Lorenzo} \sur{Briganti}}

\author[18]{\fnm{David} \sur{Brown}}

\author[3]{\fnm{Anko} \sur{B\"orner}}

\author[19]{\fnm{Cristina} \sur{Chiappini}}

\author[5]{\fnm{Cilia} \sur{Damiani}}

\author[20]{\fnm{Patrick} \sur{Gaulme}}

\author[2]{\fnm{Pablo} \sur{Heise Huijse}}

\author[6]{\fnm{Ulrike} \sur{Heiter}}

\author[21]{\fnm{Krysztof G.} \sur{Helminiak}}

\author[2]{\fnm{Mykyta} \sur{Kliapets}}

\author[22]{\fnm{Mikkel N.} \sur{Lund}}

\author[23,24]{\fnm{Paola} \sur{Marrese}}

\author[25]{\fnm{Thibault} \sur{Merle}}

\author[6]{\fnm{Nicola} \sur{Miller}}

\author[7]{\fnm{Josefina} \sur{Montalban}}

\author[26]{\fnm{Dinil Bose} \sur{Palakkatharappil}}

\author[3]{\fnm{Carsten} \sur{Paproth}}

\author[27]{\fnm{Isabel} \sur{Rebollido Vazquez}}

\author[1]{\fnm{Nena} \sur{Scheller}}

\author[28]{\fnm{Sophia} \sur{Sulis}}

\author[29]{\fnm{Amaury H. M. J.} \sur{Triaud}}

\author[19]{\fnm{Marica} \sur{Valentini}}

\author[4]{\fnm{Mathieu} \sur{Vrard}}


\author[28]{\fnm{Magali} \sur{Deleuil}}

\author[5]{\fnm{Laurent} \sur{Gizon}}

\author[30]{\fnm{Ana M.} \sur{Heras}}

\author[31]{\fnm{J. Miguel} \sur{Mas-Hesse}}

\author[32]{\fnm{Hugh} \sur{Osborn}}

\author[7]{\fnm{Isabella} \sur{Pagano}}

\author[16]{\fnm{Giampaolo} \sur{Piotto}}

\author[18]{\fnm{Don} \sur{Pollacco}}

\author[33]{\fnm{Roberto} \sur{Raggazoni}}

\author[34]{\fnm{Gavin} \sur{Ramsay}}

\author[35,36]{\fnm{Heike} \sur{Rauer}}

\author[37]{\fnm{Stephane} \sur{Udry}}

\affil*[1]{\orgdiv{Institute for Astro- and Particle Physics}, \orgname{Universit\"at Innsbruck}, \orgaddress{\street{Technikerstra{\ss}e 25}, \city{Innsbruck}, \postcode{6020}, \country{Austria}}}

\affil[2]{\orgdiv{Institute of Astronomy}, \orgname{KU Leuven}, \orgaddress{\street{Celestijnenlaan 200D}, \city{Leuven}, \postcode{3001}, \country{Belgium}}}

\affil[3]{\orgdiv{Institut f\"ur Weltraumforschung}, \orgname{Deutsches Zentrum f\"ur Luft- und Raumfahrt}, \orgaddress{\street{Rutherfordstr. 2}, \city{Berlin}, \postcode{12489}, \country{Germany}}}

\affil[4]{\orgname{Universit\'{e} C\^{o}te d'Azur, Observatoire de la C\^{o}te d'Azur, CNRS}, \orgdiv{Laboratoire Lagrange}, \orgaddress{\street{Bd de l'Observatoire, CS 34229}, \city{Nice cedex 4}, \postcode{6304}, \country{France}}}

\affil[5]{\orgdiv{Solar and Stellar Interiors Department}, \orgname{Max Planck Institute for Solar System Research}, \orgaddress{\street{Justus-von-Liebig-Weg 3}, \city{G{\"o}ttingen}, \postcode{37077}, \country{Germany}}}

\affil[6]{\orgdiv{Department of Physics and Astronomy}, \orgname{Uppsala University}, \orgaddress{\street{Box 516}, \city{Uppsala}, \postcode{75120}, \country{Sweden}}}

\affil[7]{\orgdiv{INAF-Osservatorio Astrofisico di Catania}, \orgname{Istituto Nazionale di Astrofisica (INAF)}, \orgaddress{\street{Via S. Sofia, 78}, \city{Catania}, \postcode{95123}, \country{Italy}}}

\affil[8]{\orgdiv{Astrophysics Group}, \orgname{Keele University}, \city{Keele}, \postcode{ST5 5BG}, \country{United Kingdom}}

\affil[9]{\orgdiv{Dipartimento di Fisica e Astronomia ``Augusto Righi''}, \orgname{Universit\`{a} di Bologna}, \orgaddress{\street{via P. Gobetti 93/2}, \city{Bologna}, \postcode{40129}, \country{Italy}}}

\affil[10]{\orgdiv{Space sciences, Technologies and Astrophysics Research (STAR) Institute}, \orgname{Universit\'{e} de Li\`{e}ge}, \orgaddress{\street{Quartier Agora, All\'{e}e du 6 Ao\^{u}t 19c, B\^{a}t. B5c}, \city{Li\`{e}ge}, \postcode{B4000}, \country{Belgium}}}

\affil[11]{\orgdiv{LESIA}, \orgname{Observatoire de Paris - Site de Meudon}, \orgaddress{\street{5 Place Jules Janssen}, \city{Meudon}, \postcode{92195}, \country{France}}}

\affil[12]{\orgdiv{LIRA}, \orgname{Observatoire de Paris, Universit\'{e} PSL, Sorbonne Universit\'{e}, Universit\'{e} Paris Cit\'{e}, CY Cergy Paris Universit\'{e}, CNRS,}, \orgaddress{\city{Meudon}, \postcode{92190},  \country{France}}}

\affil[13]{\orgdiv{MIT Kavli Institute for Astrophysics and Space Research}, \orgname{Massachusetts Institute of Technology}, \orgaddress{\street{70 Vassar St}, \city{Cambridge}, \postcode{02139}, \country{USA}}}

\affil[14]{\orgdiv{Departamento de Astrof\'{i}sica}, \orgname{Universidad de la Laguna}, \orgaddress{\city{La Laguna, Tenerife}, \postcode{E-38200}, \country{Spain}}}
\affil[15]{\orgdiv{Instituto de Astrof\'{i}sica de Canarias}, \orgaddress{\city{La Laguna, Tenerife}, \postcode{E-38200}, \country{Spain}}}

\affil[16]{\orgdiv{Dipartimento di Fisica e Astronomia}, \orgname{Universit\`{a} di Padova}, \orgaddress{\street{Vicolo dell'osservatorio 3}, \city{Padova}, \postcode{35122}, \country{Italy}}}

\affil[17]{\orgdiv{INAF - Osservatorio di Astrofisica e Scienza dello Spazio, Bologna}, \orgaddress{\street{via P. Gobetti 93/3}, \city{Bologna}, \postcode{40129}, \country{Italy}}}

\affil[18]{\orgdiv{Department of Physics}, \orgname{University of Warwick}, \orgaddress{\street{Gibbet Hill Road}, \city{Coventry}, \postcode{CV4 7AL}, \country{Country}}}

\affil[19]{\orgname{Leibniz Institute for Astrophysics Potsdam}, \orgaddress{\street{An der Sternwarte 16}, \city{Potsdam}, \postcode{14482}, \country{Germany}}}

\affil[20]{\orgname{Th\"uringer Landessternwarte Tautenburg}, \orgaddress{\street{Sternwarte 5}, \city{Tautenburg}, \postcode{07778}, \country{Germany}}}

\affil[21]{\orgdiv{Nicolaus Copernicus Astronomical Center}, \orgname{Polish Academy of Sciences}, \orgaddress{\street{ul. Rabianska 8}, \city{Tor\'{u}n}, \postcode{87-100}, \country{Poland}}}

\affil[22]{\orgdiv{Department of Physics and Astronomy}, \orgname{Aarhus University}, \orgaddress{\street{Ny Munkegade 120}, \city{Aarhus}, \postcode{8000}, \country{Denmark}}}

\affil[23]{\orgdiv{Space Science Data Center}, \orgname{Italian Space Agency}, \orgaddress{\street{Via del Politecnico, snc, Edificio D}, \city{Rome}, \postcode{00133}, \country{Italy}}}

\affil[24]{\orgdiv{INAF - Italian National Institute for Astrophysics}, \orgname{Astronomical Observatory of Rome}, \orgaddress{\city{Rome},  \country{Italy}}}

\affil[25]{\orgdiv{Royal Observatory of Belgium}, \orgname{Universit\'{e} Libre de Bruxelles}, \city{Brussels}, \country{Belgium}}

\affil[26]{\orgdiv{D\'{e}partement d'Astrophysique}, \orgname{CEA Paris Saclay}, \orgaddress{\street{L'Orme des Merisiers, bat. 709}, \city{Gif-sur-Yvette}, \postcode{91191}, \country{France}}}

\affil[27]{\orgdiv{European Space Astronomy Centre (ESAC)}, \orgname{European Space Agency (ESA)}, \orgaddress{\street{Camino Bajo del Castillo s/n}, \city{Villanueva de la Ca\~{n}ada}, \postcode{28692}, \country{Spain}}}

\affil[28]{\orgdiv{Laboratoire d'astrophysique de Marseille (LAM)}, \orgname{Universit\'{e} Aix Marseille, CNRS, CNES}, \orgaddress{\street{38 Rue Fr\'{e}d\'{e}ric Joliot Curie}, \city{Marseille}, \postcode{13013}, \country{France}}}

\affil[29]{\orgdiv{School of Physics and Astronomy}, \orgname{University of Birmingham}, \orgaddress{\street{Edgbaston}, \city{Birmingham}, \postcode{B15 2TT}, \country{United Kingdom}}}

\affil[30]{\orgdiv{European Space Agency (ESA)}, \orgname{European Space Research and Technology Centre (ESTEC)}, \orgaddress{\street{Keplerlaan 1}, \city{Noordwijk}, \postcode{2201 AZ}, \country{The Netherlands}}}

\affil[31]{\orgname{Centro de Astrobiolog\'{i}a (CAB) CSIC-INTA}, \orgaddress{\street{Camino Bajo del Castillo s/n}, \city{Villanueva de la Ca\~{n}ada, Madrid}, \postcode{28692}, \country{Spain}}}

\affil[32]{\orgdiv{Physikalisches Institut}, \orgname{University of Bern}, \orgaddress{\street{Gesellschaftsstrasse 6}, \city{Bern}, \postcode{3012}, \country{Switzerland}}}

\affil[33]{\orgdiv{INAF-Osservatorio Astronomico di Padova}, \orgaddress{\street{Vicolo dell' Osservatorio 5}, \city{Padova}, \postcode{35122}, \country{Italy}}}

\affil[34]{\orgdiv{Armagh Observatory and Planetarium}, \orgaddress{\street{College Hill}, \city{Armagh}, \postcode{BT61 9DG}, \country{Northern Ireland, UK}}}

\affil[35]{\orgdiv{Deutsches Zentrum f\"ur Luft- und Raumfahrt}, \orgaddress{\street{Markgrafenstraße 37}, \city{Berlin}, \postcode{10117}, \country{Germany}}}
\affil[36]{\orgdiv{Institut f\"ur Geologische Wissenschaften}, \orgname{Freie Universit\"at Berlin} \orgaddress{\street{Malteserstraße 74-100}, \city{Berlin}, \postcode{12249}, \country{Germany}}}

\affil[37]{\orgdiv{Observatoire de Gen\`{e}ve}, \orgaddress{\street{51 chemin des Maillettes}, \city{Sauverny}, \postcode{1290}, \country{Switzerland}}}


\abstract{In order to meet the science goals of the PLATO space mission, an
extensive science calibration and validation plan has been
designed. This paper describes this plan, as well as the methodology
adopted to select the science calibration and validation stars that
have entered its input catalogue. This is the so-called {\tt scvPIC}, which
is part of the general PLATO Input Catalogue (PIC) for the first
selected long pointing field in the Southern Hemisphere known as
LOPS2.  While many of PLATO's science requirements needed dedicated
stars as calibrators as discussed here, its most stringent requirement
is the delivery of the age of the host stars of exoplanetary systems
with an accuracy better than 10\% for a G0V star of {\it V} = 10 mag, i.e. a nearby Sun-like star. This is presently not within reach
for large populations of dwarfs and subgiants in the Milky Way as it
requires the models of their stellar interiors to be improved. We
discuss how this ambitious age requirement led to the selection of
tens of thousands of red giants, and of thousands of main-sequence
early F-type gravity-mode pulsators in order to deduce their internal
rotation profile across stellar evolution. This asteroseismic
observable will then be imported as key information into improved
models of dwarfs and subgiants in the Milky Way as optimal modelling tools for ever better age-dating of the exoplanet hosts as the PLATO mission moves
along. Additional calibrators and validators included in the {\tt scvPIC} are a few thousands of binaries, a few hundreds of legacy and benchmark stars, a few hundred photometrically stable stars, and six transiting brown dwarfs.}




\maketitle

\section{Introduction}\label{sec1}

The PLATO mission (PLAnetary Transits and Oscillations of stars) has been selected as part of ESA’s Cosmic Vision 2015–2025 program for the M3 mission launch currently foreseen in early 2027. The main science goal of PLATO is to detect and characterise extrasolar planets, including terrestrial planets in the habitable zone (HZ) of their solar-type host stars \citep{Rauer2025}. 
Characterisation here means to derive {\it accurate\/} planetary radii, masses, and ages. As transit searches offer only an indirect detection and characterisation method, relative to the host star, the accurate knowledge of the host star parameters drives the accuracy of the derived planet parameters. PLATO will monitor about 150,000 stars per pointing for two or more years with the aim to maximally characterise these stars using asteroseismology and to detect transiting planets. 

PLATO also has a complementary science program (PLATO-CS) open to the world by means of a Guest Observer (GO) call for proposals
\citep{Rauer2025}. The GO calls to be issued by the European Space Agency (ESA) are planned about nine months before the start of each long pointing, with the first call opening on 7 April 2026\footnote{https://www.cosmos.esa.int/web/plato/getting-ready}.  Eligible GO applications may cover any science topic outside the core science of the mission, as long as it can be achieved from objects within the nominal field of view for which the call occurs.
The resulting large data set of light curves will hence provide numerous additional science returns of the mission reaching far beyond just the exoplanet science case, notably stellar, galactic, and extragalactic astrophysics \citep{Tkachenko2024}. 

The PLATO payload has a multi-telescope configuration consisting of 26 cameras, each with an aperture of 12 cm \citep{Pertenais2021, Rauer2025}. These cameras cover a total field of view of 
2132 square degrees covered by 104 CCDs. Altogether PLATO's CCDs consist of 2000 Megapixels \citep{Pertenais2021}.  
Twenty four of these cameras (i.e., the normal cameras or N-CAMs) will operate in white light providing photometric time series of tens of thousands of objects. For thousands of stars, 6-pixel by 6-pixel images (so-called ``imagettes'') will be downloaded with cadences of 25 seconds, while fainter objects will receive 50-seconds or 10-minute cadence onboard computed light curves \citep{Rauer2025}. The two remaining cameras (i.e., the fast cameras or F-CAMs) carry a blue (505-700 nm) and a red (665-1000 nm) filter, respectively, and will provide colour information by means of 2.5-second cadence 6-pixel by 6-pixel imagettes for selected bright targets in the visual magnitude range from four to eight. More details on the performance of the instrument are provided in Cabrera et al. (2026, submitted to {\it Experimental Astronomy}). 

\begin{figure}[htb]
\centering
\includegraphics[width=0.75\textwidth]{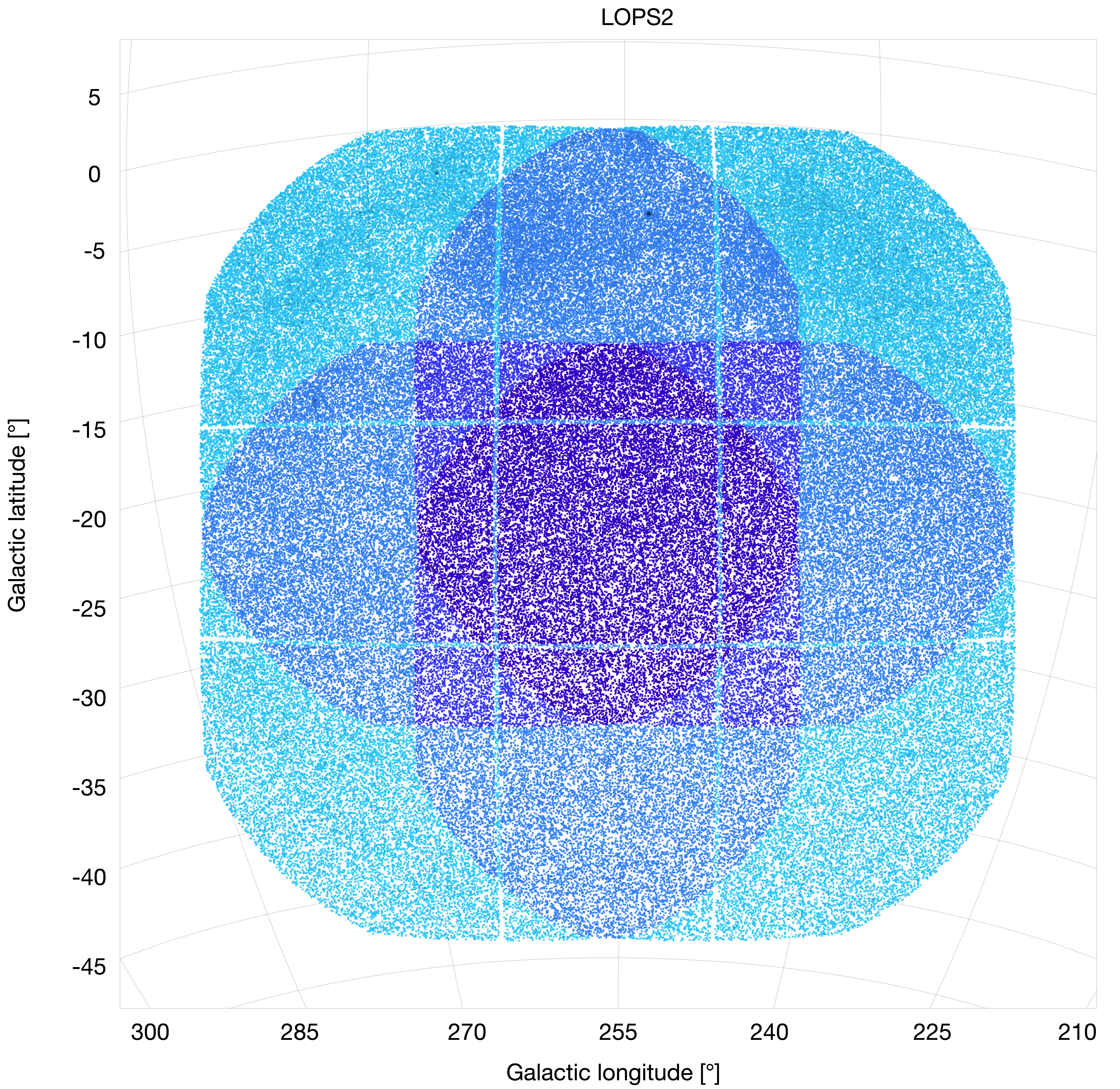}
\caption{Location of LOPS2 based on PIC v2.2.0.1 in galactic coordinates. Light, medium, darker, and darkest blue colours show the areas monitored by 6, 12, 18 and 24 normal cameras (N-CAMs), respectively. The blue points are all targets in the PIC making up the `PLATO footprint'.}
\label{fig:LOPS2}
\end{figure}

The nominally funded science operation phase of PLATO is currently planned for four years \citep[][onboard consumables available for 8.5 years]{Rauer2025} with so-called Long-duration Observing Phases (LOP) of either twice two years 
or observations of only one LOP field for the full duration of the mission \citep{Rauer2025}. The final choice of operation mode will only be made once in orbit, so as to rely on the most up to date knowledge of exoplanet science at that time.
Two long pointing target fields have been selected: one field in the Northern hemisphere (LOPN1) and one field in the Southern hemisphere (LOPS2). Based on the availability of ground-based follow-up observations, the first field to be observed by PLATO will be LOPS2 in the Southern Hemisphere (see Figure \ref{fig:LOPS2}). The selection process for these fields and the characterisation of the LOPS2 can be found in \cite{Nascimbeni2025}.
The LOPS2 field\footnote{An animated version of LOPS2 using PIC v2.2.0.1 can be found on YouTube under \href{https://youtu.be/rU3fas8NFHY}{https://youtu.be/rU3fas8NFHY}} is centred on the coordinates $\alpha\,=\,06^h\,:21^m\,:14.5^s$ and $\delta=-47^{\circ}:53^{\prime}:13^{\prime\prime}$ ($l=255.9375^{\circ}, b=-24.62432^{\circ}$).

To reach the core science objectives of PLATO, four stellar samples were defined: P1, P2, P4, and P5\footnote{P3 was dropped during the development of the mission.} \citep{Rauer2025}.
The central targets for PLATO are comprised in the bright samples P1 and P2. In P1 at least 15,000 dwarf and sub-giant stars with spectral types from F5 to K7 and $V$ magnitudes brighter than 11 mag are included considering the two pointings altogether. The 1,000 targets of sample P2 have the same spectral types as the P1 sample targets, but are brighter than $V$ magnitude 8.5. Hence, P2 targets are a sub-sample of the P1 sample. 
The P4 sample includes at least 5000 cool late-type dwarfs with spectral types from late K to M that are brighter than $V = 16$ mag in the solar vicinity. The statistical sample P5 contains at least 245,000 dwarf and subgiant stars with spectral types F5 to K7 that are brighter than $V = 13$ mag. Details about the PLATO stellar samples can be found in \cite{Rauer2025}.

Several data products will be available from the PLATO observations. Level-0 (L0) data are unprocessed data that are downloaded for the targets from all cameras. Level-1 (L1) data are calibrated data including, for example, corrections for instrumental effects. L1 data are produced on-ground for all targets and all cameras. Level-2 (L2) data are the first scientific data products that include scientific information such as transit candidates and their parameters on the exoplanet core science side and results of asteroseismic analyses on the stellar core science side. Level-3 (L3) data comprise the final PLATO Catalogue, and Lg data include results of the PLATO ground-based radial-velocity follow-up campaigns. A more detailed description of the PLATO data products can be found in \cite{Rauer2025}.

As already mentioned, PLATO will also provide high-precision photometric time series for thousands of targets that are of other types than the core science targets. All science that addresses topics other than the core science is included in PLATO-CS in order to maximize the scientific return of the mission \citep{Rauer2025}. The catalogue of science calibration and validation targets (i.e., the {\tt scvPIC}) bridges core and complementary science and is needed to reach the main scientific goals of the mission.


\section{Science calibration and validation plan for PLATO}

The detection and precise characterisation of the PLATO stellar and planetary targets requires excellent photometric measurements. 
Additionally to achieve an excellent performance of the instruments, calibration and characterisation activities are implemented for the PLATO mission at different levels. These can be divided into instrument and science calibration.

\subsection{Instrument calibration}

Before launch, calibration and characterisation activities concentrate on the instrument performance at unit and camera level. 
The difference between {\it calibration\/} and {\it characterisation\/} comes from the difference in use of the measurements. In the case of characterisation, the measurements are stored as information that can be used in the future, but they have no influence on further tests. For example, CCD dark current can be characterised as a function of temperature to understand how the shot noise contribution depends on the CCD temperature. On the other hand, a calibration allows for the determination of a new model description or a transfer function between two characteristics of the system. As an example, the relation between a temperature measurement and the telescope focus position can be characterised to derive a calibration function, which can influence other tests or can be used for science operations. More details on the performance of the instrument can be found in Cabrera et al. (2026; to be submitted to {\it Experimental Astronomy}). 

In flight, the calibration and characterisation activities concentrate on mission performance. The in-orbit verification supplements/confirms ground verification by providing operating conditions which cannot be fully or cost effectively duplicated or simulated on ground. It characterises the system under operational conditions, especially for the aspects that cannot be determined before the launch. The in-orbit verification ideally confirms that the space and ground elements are compatible with each other. It is divided into two parts: (i) the calibration during the commissioning phase, where the most relevant parameters are set, such as the operating temperature for each camera obtained from tuning of the focus; and (ii) the calibration during nominal science operations, where the in-orbit evolution of
the instrument is monitored (ageing) and regular characterisation activities are performed (background monitoring, CCD bias monitoring, etc.). 

\subsection{Science calibration and validation}
Aside from the instrument calibration, PLATO core science requires a dedicated and extensive scientific calibration plan, notably to derive and deliver the L2, L3, and Lg data products. 
L1 and L2 pipelines are scientifically validated using FGK dwarf and sub-giants oscillations and planetary transits.
The derivation of the high-level products (i.e., L2, L3, and Lg data) involves the use of algorithms relying on theoretical models of stellar interiors and of stellar atmospheres. This is where asteroseismic modelling based on detected and identified oscillation modes comes in \citep[e.g.,][for a general review on how such modelling works]{Aerts2021}. The analysis pipelines rely on grids of stellar structure and evolution models. These models, while quite sophisticated already, still need improvements to meet the 10\% accurate age requirement, notably on the front of transport processes \citep{Mathis2013,Aerts2019}.
At present, the available stellar models are not yet on par with the scientific requirements for large samples of dwarfs and subgiants. 

The PLATO Science Calibration and Validation plan serves the purpose to upgrade the quality of the models of stellar interiors, from asteroseismology of dedicated calibration stars in PLATO's fields, starting with LOPS2. In the rest of this paper we therefore focus on the 
target selection to achieve optimal science calibration, with attention to the most stringent of the science requirements, namely a 10\% age accuracy of the dwarfs and subgiants in the prime sample of the mission. In order to reach this, it is necessary to better understand the internal rotation profiles of such stars, as rotation induces a myriad of instabilities in the stellar interior. 

We present the science calibration and validation plan to be implemented during nominal science operations. These are calibration activities during mission science operations required to guarantee the success of the mission on the front of stellar characterisation (e.g.\ identification and characterisation of oscillation mode frequencies of stars) and planetary characterisation (e.g.\ unbiased radius determination in the presence of the stellar variability signal).
For science calibration and validation purposes, six different samples of targets were selected: binaries ({\tt scv1}), legacy and benchmark stars ({\tt scv2}), photometrically stable stars ({\tt scv3}), red giants ({\tt scv4}), $\gamma$ Doradus stars ({\tt scv5}), and known transiting brown dwarfs ({\tt scv6}). 


\section{The Science Calibration and Validation plan and its targets for LOPS2}
\label{sec:scvtargets}

While nearly 8000 exoplanets are known to date, among which more than 4600 transiting their host star\footnote{The numbers of exoplanets were derived in Feb. 2026 from \url{https://exoplanet.eu/home/.}}, only about a hundred exoplanet host stars have been age-dated with high precision between 10\% and 20\% from asteroseismology. Moreover, the majority of the host stars with such level of age-dating is fainter than $V$ magnitude 11.  High-precision asteroseismic ages have mainly been achieved for rather faint exoplanet host stars 
\citep[][their Fig.\,8]{Chontos2021}. 
Currently, asteroseismic age-dating reaches about 10\% precision for a few tens of relatively bright sun-like dwarfs whose metallicity, surface abundances, effective temperature, gravity, and surface rotation have been measured to better than $\sim\,$10\% precision. This result relies on the adoption of internal and atmospheric physics calibrated by helioseismology \citep{Lebreton2014,Chaplin2014,Lund2017,SilvaAguirre2017,Gent2022} and therefore concerns precision with respect to solar-scaled models. No star is sufficiently similar to the Sun to avoid biases in the stellar age-dating stemming from assumed fixed input physics. Such biases result in systematic uncertainties in addition to statistical uncertainties in the parameter estimation from stellar modelling, including the mass, radius, and age. 

The PLATO core science aims for an accuracy of 2\% in stellar radius, 4\% in stellar mass and 10\% in age for a G0\,V star with $m_{\rm V} \leq 10$ mag \citep{Rauer2025}. 
Of these requirements, the one on the age is by far the most demanding because it is strongly dependent on the chosen input physics assumed in the stellar models \citep[e.g.,][]{Chaplin2014,SilvaAguirre2017}.  In fact, if the age requirement is fulfilled for the exoplanet host stars in the P1 and P2 samples, which constitute the backbone of PLATO, the other listed requirements are also met. For this reason, the bulk of the science calibration and validation plan focuses on the 10\% age requirement for the dwarfs and subgiants with spectral types from F5 to K8 with magnitudes below $m_V=11$\,mag (for the normal cameras) and between 4\,mag and 8\,mag (for the fast cameras).

The PLATO core science needs proper scientific calibration. This will be achieved by the monitoring of thousands of stars. This subset of the PLATO Input Catalogue (PIC) is known as the science calibration and validation sample ({\tt scvPIC}).
As will become clear throughout the following text, many of the PLATO science
calibration or validation stars fall under the mission’s core program (i.e., are P1 stars), while others do not. That is because some aspects of the science calibration, notably to reach 10\%-level stellar age-dating, cannot be done from currently available data. PLATO will be the first infrastructure to deliver data to achieve some aspects of the scientific calibration of stellar models, and hence of the mission. As such, all the data of the targets in the {\tt scvPIC} discussed below and belonging to the core program will only become public following the nominal PLATO data releases. 
On the other hand, all the L0/L1 data products of the scvPIC stars that are not part of the core program will become public to the worldwide community with a timeline yet to be finalised.
Table \ref{tablesummary} gives an overview of the requested science calibration and validation stars and the way they should be observed for optimal effect and capacity.

\begingroup

\setlength{\tabcolsep}{2.5pt} 
\renewcommand{\arraystretch}{1.5} 

\begin{table}[htbp]
  \caption{Summary of requested science calibration and validation stars.
  The upper part of the table lists the calibrators and validators for stellar science, the middle part for exoplanetary science, and the lower part list stars for timing calibration purposes. }
\label{tablesummary}
\centering
\begin{footnotesize}
\begin{tabular}{cccc}
\toprule
Science calibration and validation targets  & Type	& Number of stars &  In core program?\\
\midrule
DEB@4\%mass accuracy, SpT from A9--M9 & V &	$\leq 200$  & most of them \\
Interferometry@1\% radius precision, SpT F--M & V &	$\leq 500$  & yes, all \\
TESS legacy stars  & V	& $\leq 100$  & yes, all \\
Gaia benchmark stars/PLATO validators & V		& $\leq 250$  & most of them \\
$\Omega(r,t), D_{\rm mix}(r,t)$ probes: dwarfs, SpT F5--K8 & C &	 $800$  & yes, all \\
$\Omega(r,t), D_{\rm mix}(r,t)$ probes: subgiants  & C &	 $300$  & yes, all \\
$\Omega(r,t), D_{\rm mix}(r,t)$ probes: red giants  & C &	 $24000$  & none \\
$\Omega(r,t), D_{\rm mix}(r,t)$ probes: $\gamma\,$Dor stars & C	& $1440$  & none\\
\midrule
Known exoplanets or DEB, $m_V\leq 11$ & C,V & 	$\leq 200$  & exo yes, EB no \\
Known exoplanets or DEB, $11\leq m_V\leq 13$  & C,V &	 $\leq 2000$  & exo  yes, EB  no \\
\midrule
Ap/Bp stars, monoperiodic or constant & T &	$\leq 50$  & none\\
@49\,ppm/hr, $m_V\leq 11$ & &	& \\
Ap/Bp stars, monoperiodic or constant  & T & $\leq 50$   & none\\
@49\,ppm/hr, $11\leq m_V\leq 13$ & &	& \\
\botrule
\end{tabular}
\end{footnotesize}
\footnotetext{The types of targets are calibrators (C), validators (V) and stars for timing calibration (T) purposes. DEB stands for detached eclipsing binary and SpT for spectral type, 
The evolution of the internal rotation profile in time is abbreviated as $\Omega(r,t)$, and the overall internal mixing profile as $D_{\rm mix}(r,t)$. }
\end{table}

\endgroup

In the following subsections we describe the background for the six different calibration and validation samples and how the respective targets were selected. For most of the {\tt scvPIC} targets, data from the TESS mission \citep[Transiting Exoplanet Survey Satellite;][]{Ricker2015} assisted the target selection. 
For all targets within the {\tt scvPIC} we used the following priority metric: we assigned priority 1 for targets that are crucial to achieve the calibration and validation goals for each subsample; targets with priority 2 extend the parameter space for calibration purposes; all other targets that fit the {\tt scv} sample definition have priority 3.

\subsection{Binaries ({\tt scv1})} 

The {\tt scv1} catalogue includes different types of binary stars, and, hence is divided into five sub-samples: eclipsing binaries ({\tt scv1a}), astrometric binaries ({\tt scv1b}), wide binaries ({\tt scv1c}), HW Vir-type binaries ({\tt scv1d}), and wide white dwarf binaries ({\tt scv1e}).
Binary stars are an essential part of the science calibration activities because of the possibility to measure precise model-independent masses for spectroscopic binary systems where the inclination is known from the analysis of the light curve for eclipsing binary systems ({\tt scv1a}) or from the astrometric orbit for astrometric binary systems ({\tt scv1b}). 
Detached eclipsing binary stars also offer the possibility to measure precise stellar radii. 
This can be a very powerful means to calibrate stellar models, particularly if combined with asteroseismology
\citep{SouthworthBowman2026}. 
The availability of precise parallaxes for detached eclipsing binary stars from the Gaia mission also makes it possible to directly measure the effective temperatures of the stars in the these systems from the angular diameters and bolometric fluxes \citep{2020MNRAS.497.2899M}.

TESS light curves are available for all the eclipsing binary stars included in the {\tt scv1} catalogue. 
PLATO light curves will be observed in a different passband to TESS so it will be possible to check that the radii measured are accurate by checking the consistency of the results obtained with both instruments. 
Eclipsing binary stars also give the opportunity to directly measure the limb-darkening of stars.

Wide binary systems ({\tt scv1c}) are included as a way to calibrate methods to estimate stellar ages against the asteroseismic age scale, and to validate the consistency of the age estimates for different methods and different types of stars. Subsample {\tt scv1d} consists of a single star (AA Dor) which is an eclipsing binary star suitable for end-to-end tests of the time stamps provided with the PLATO data products.

There are about 50 stars from the core P1 sample (``P1 stars'') in the LOPS2 field that have a distant white dwarf (WD) companion. These wide binary systems can be identified from the common proper motion (CPM) of the two stars. Observations of these WD-CPM binaries ({\tt scv1e}) will be essential to validate the age estimates from asteroseismology for the P1 stars, particularly for old stars where there will be no suitable validation targets obtained from stars in open clusters. The best age estimates come from massive WD because there is then little uncertainty in the main-sequence lifetime of the progenitor inferred from the initial-final mass relation \citep{Rebassa2021}. Those massive WDs where the age estimate is of comparable precision to that estimate from asteroseismology are an important contribution to the science calibration effort. These binaries give us the opportunity to optimise the free parameters in the new physics in the stellar models by making the age estimates from asteroseismology and white dwarf cooling models consistent with one another.

For the selection of binary stars that might be observable in the first two years of the PLATO mission, we defined the edge of the nominal LOPS2 field using the galactic coordinates of all stars listed in the PLATO input catalogue version PIC2.0.0.1-t. 
We computed the position angle $\theta_{\rm LOPS2}$ and angular separation on the sky $r_{\rm LOPS2}$ for all stars relative to the centre of the field, and then took the maximum value of $r_{\rm LOPS2}$ in 180 bins of width 2$^{\circ}$ in $\theta_{\rm LOPS2}$ to define the edge of the nominal LOPS2 field.  
We then extended the maximum value of $r_{\rm LOPS2}$ in each bin by 6\% to define the selection boundary. 
The additional 6\% allows for some tolerance in the absolute pointing of the spacecraft and the usable area of the images produced by the N-CAMs.

\subsubsection{{\tt scv1a}: eclipsing binaries}
Several catalogues were used to select the {\tt scv1a} sample, each of which discussed here.
DEBCat is a catalogue of the physical properties of well-studied detached eclipsing binary stars \citep[DEBs;][]{Southworth2015}. Eclipsing binaries are selected with mass and radius measurements accurate to 2\%, plus a few interesting systems which do not quite meet this criterion. The catalogue contains binaries which are representative of single stars: ones whose evolution was/is affected by mass transfer are not included. It is also required for both stars to have a measurement of their effective temperature. The “live” catalogue\footnote{\url{https://www.astro.keele.ac.uk/jkt/debcat/}} is maintained and regularly updated by researchers from Keele University. Targets were selected using the version of the catalogue downloaded 2024 April 3. Stars fainter than $V=13.5$ mag, as listed in the catalogue, were excluded. There are 12 systems within the LOPS2 field.

\cite{Maxted2023a} described the selection and initial characterisation of 20 eclipsing binary stars that are suitable for calibration and testing of stellar models and data analysis algorithms used by the PLATO mission and spectroscopic surveys. The binary stars selected are F-/G-type dwarf stars with M-type dwarf companions that contribute less than 2\% of the flux at optical wavelengths. Of the four stars within LOPS2, three are P1 stars that can be used for end-to-end tests of the stellar parameters derived from asteroseismology if solar-like oscillations are detected. 

\begin{figure}
\includegraphics[width=0.95\textwidth]{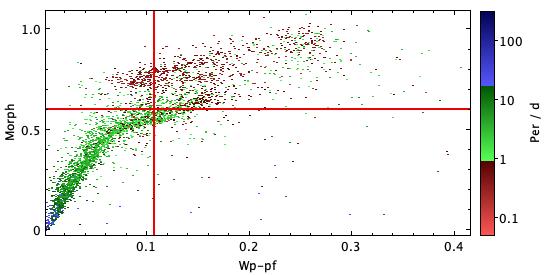}
\caption{Light curve morphology parameter (Morph) as a function of the width of the primary eclipse in phase units (Wp-pf) for eclipsing binary stars from Pr{\v{s}}a et~al. (2022). Points are colour-coded by the reported orbital period of the binary in days (Per / d). The horizontal and vertical lines define the region used to select targets for subsample {\tt scv1a}.}
\label{fig:tessebs_morph_pwidth}
\end{figure}

\cite{Prsa2022} have published a catalogue of 4584 eclipsing binaries observed during the first two years (26 sectors) of the TESS survey. This catalogue includes a morphology parameter, defined as a continuous variable in the range $[0,1]$, where 0 corresponds to the widest detached systems and 1 to ellipsoidal variables. The value of this parameter is determined using a neural-network model trained on the {\it Kepler} EB data set. Fig.~\ref{fig:tessebs_morph_pwidth} shows the value of this parameter (Morph) as a function of the primary eclipse width in phase units measured using a high-order polynomial fit to the phase-folded light curve (Wp-pf). Points are colour-coded according to the orbital period from the same catalogue. As expected, EBs with large morphological parameters (Morph $\gtrsim 0.60$) are typically short-period systems (P $\lesssim 1$ day). The stars in these binary systems will be strongly distorted by their mutual gravitational potential and so are not suitable for testing models of single stars.

One of the requirements for {\tt scv1a} is that all late-type targets (with spectral type F5V or later) shall have variability between the eclipses with an amplitude of less than 1\%. This requirement was designed to avoid the inclusion of these highly-distorted binary stars, particularly RS CVn binaries where the rapid rotation of the stars leads to extremely high levels of magnetic activity. It is not straightforward to calculate the variability between the eclipses for a large sample of stars automatically, so we use the Morph parameter as a proxy for this parameter. As a guideline, we used {\sc jktebop}\footnote{\url{https://www.astro.keele.ac.uk/jkt/codes/jktebop.html}} \citep{jktebop} to compute light curves for two identical solar-type stars in a circular orbit for various values of their fractional radius, $r= R_{\star}/a$, where a
is the semi-major axis of the orbit. We found that these model light curves have a peak-to-peak amplitude outside
eclipse of 1\% when the sum fractional radii is $r_1 + r_2 = 0.33$. The oblateness of the stars in such a binary is approximately 0.7\% and the width of the primary eclipse is 0.107 phase units. This limit is shown in Fig.~\ref{fig:tessebs_morph_pwidth} where it can be seen that it corresponds to a limit Morph $\approx 0.6$ if we ignore short-period systems. We therefore decided to exclude systems with Morph $\geq 0.6$ from the selection of stars for sub-sample {\tt scv1a} from this catalogue. We also excluded stars with $V$ magnitudes $V \geq 13$ listed in the TIC.

There are 341 stars listed in \cite{Prsa2022} within LOPS2 that satisfy the selection criteria Morph $< 0.6$ and $V < 13$. Of these, 61 are P1 stars. There are also 11 stars listed as planet-host stars. One of these is the circumbinary planet system TOI-1338 \citep{Sebastian2024}, the others are transiting exoplanets that have been mis-classified as EBs.

EBLM systems are solar-type stars in EBs with very low-mass companions \citep{Maxted2023}. Stars with a mass $M \lesssim 0.35 M_{\odot}$, are approximately the same size as the largest hot Jupiters so the transit of a solar-type star by either a hot Jupiter or a very low-mass star will produce a dip in the light curve with a depth of about 1\%. EBLM systems can be used to measure limb-darkening using similar methods to those already applied to hot Jupiters \citep{Maxted2023b,2025MNRAS.537.3943F}. They are also known to host transiting circumbinary planets \citep{Sebastian2024}.

\cite{Maxted2023} provide a catalogue of 138 EBLM systems that have published mass and radius estimates
for the very low-mass companion star. Of the stars brighter than $V = 13$ mag, eight lie within LOPS2 and one is a P1 star. This catalogue has been supplemented with a list of EBLM systems identified using light curves from the WASP project \citep{Pollaco2006} that are currently being followed-up by the BEBOP radial-velocity survey for circumbinary planets \citep[][]{Martin2019}. This list provides an additional 28 stars brighter than $V = 13$; 2 of which are P1 stars.

On the massive star end, \cite{Eze2024} have published a catalogue of 43 $\beta$ Cep pulsators in eclipsing binaries observed with TESS. Of these, four are brighter than $V = 13$ mag and lie within the selection boundary of LOPS2. 

CR\'EME is an observational project of high-resolution spectroscopic observations, intended to derive precise radial velocity (RV) measurements and atmospheric parameters \citep{Helminiak2022}. The main goals of the project are: (i) the identification of new examples of rare, poorly studied, or otherwise interesting DEBs. (ii) the precise characterisation of the studied systems i.e. determination of masses, radii, temperatures, distances, metallicities, and ages of stars. The project goals were defined and the observations started in 2011, and initially targeted a sample of southern hemisphere DEBs, identified by the All-Sky Automated Survey \citep[ASAS;][]{Pojmanski2002}. There are eleven DEBs brighter than $V = 13$ within the LOPS2 selection boundary from the CR\'EME project target list. Of these, three are P1 stars.

The WASP project\footnote{\url{https://wasp-planets.net}} has obtained over 580 billion photometric observations for more than 30 million bright stars ($8 \lesssim V \lesssim 13$) during a survey that has discovered almost 200 transiting exoplanets since observations started in May 2004 \citep{Pollaco2006}. Thousands of eclipsing binary stars have been identified and flagged in the WASP archive database by visual inspection of the light curves by members of the WASP project. 10 DEBs with well-defined eclipses and little out-of-eclipse variability, 9 of which are P1 stars, were selected from the list of WASP EBs after confirmation of their suitability by  visual inspection of their TESS light curves. 

\cite{Shi2022} have published a catalogue of 57 new pulsating stars in EA-type binary stars based on the analysis of TESS 2-minute cadence data from sectors 1 -- 45. Note that EA-type in this context refers to the classification of the light curve shape as listed in the International Variable Star Index \citep[VSX;][]{Watson2006} from which \cite{Shi2022} drew their initial sample of stars for analysis. These are typically Algol-type binary stars that are the outcome of substantial mass transfer between the stars. These Algol-type stars are not suitable for calibrating models of single stars. However, Table 2 of \cite{Shi2022} lists stars that are not Algol-type binary stars. Of these, four stars lie within LOPS2 and are brighter than $V = 13$ mag. Visual inspection of the light curves for these stars plotted in Figs. 1 and 2 of Shi et al. show that these are all detached eclipsing binary stars (DEBs). 

\cite{Howard2022} identified 370 EBs from approximately 510,000 short-cadence TESS light curves from sector 1 -- 26 that were not included in the catalogue published by \cite{Prsa2022}. Of these, 20 stars are brighter than $V = 13$ mag and lie within the selection boundary of LOPS2. We inspected the TESS 120-s cadence light curves for these 20 stars to verify that they are DEBs and that the amplitude of the variability between the eclipses due to effects other than pulsations (ellipsoidal effect or magnetic activity) is less than 1\%. We excluded one known transiting extrasolar planet (WASP-121). Of the remaining 19 stars, 8 are P1 stars. Pulsations are apparent from the visual inspection of the light curves for 4 of the stars
not in the P1 sample. This sample includes some DEBs with long-period binaries where it has not (yet) been possible to determine the correct orbital period from the TESS photometry alone.

\cite{Justesen2021} have analysed a total of 247,565 single-sector short-cadence TESS light curves from sectors 1–13 to produce a list of 748 eclipsing binary candidates. Of these, 134 stars were not already selected  from any of the catalogues listed above, are brighter than $V = 13$ mag, and lie within the LOPS2 selection boundary. We generated plots of the phase-folded TESS light curves and inspected these plots to identify and remove 67 stars from the sample that are clearly not DEBs or that show variability between the eclipses due the ellipsoidal effect or magnetic activity with an amplitude greater than 1\%. We also noted 9 stars where variability due to pulsations was apparent in the light curve. The star TIC 404804908 is a single entry in the TESS input catalogue that corresponds to two entries in Gaia DR3 \citep{2016A&A...595A...1G, 2023A&A...674A...1G} for two stars of similar brightness. We have selected the brighter Gaia source for inclusion in sub-sample {\tt scv1a}. Of the 67 selected stars, 7 are P1 stars.

The Gaia DR3 catalogue includes a table with the detailed variability results for 2,184,477 EB candidates. This table is dominated by faint, short-period systems that are not suitable for selection for sub-sample {\tt scv1a}. We used Vizier\footnote{\url{https://vizier.cds.unistra.fr/}} to obtain a list of 5425 EBs within $30^{\circ}$ of the centre of the LOPS2 field, with a $G$-band magnitude \texttt{geom\_model\_reference\_level}, smaller than 14, and a primary eclipse phase duration, \texttt{DurE1}, shorter than 0.1 days, where the upper limit on \texttt{DurE1} has been selected to avoid the large numbers of W UMa-type binaries in the Gaia DR3 catalogue.

We used {\sc topcat}\footnote{\url{https://www.star.bris.ac.uk/~mbt/topcat/}} \citep{topcat} to cross-match these sources with the TESS input catalogue (TIC) version v8.2 \citep{TICv82}  to select 2292 stars brighter than $V = 13$ mag. We find that 1955 of these stars lie within the selection boundary of LOPS2 and 1884 are not among the stars selected from the catalogues noted above. We used visual inspection of the TESS light curves and Gaia DR3 epoch photometry for these 1884 stars to identify and measure the orbital periods of 329 detached eclipsing binary stars with variability between the eclipses due to effects other than pulsations (ellipsoidal effect or magnetic activity) with an amplitude less than 1\%. Of these 329 DEBs, 4 are P1 stars. Many of the stars selected from this catalogue are found at low galactic latitude – 223 out of 309 stars are found at $|b| < 10^{\circ}$.

The  American Association of Variable Star Observers (AAVSO) maintain a database of variable stars (The International Variable Star Index) that is accessible through the VSX website\footnote{\url{https://www.aavso.org/vsx/index.php}}. We downloaded the data for 4674 EA-type variable stars within $30^{\circ}$ of the centre of the LOPS2 field on 2024 April 09. We used {\sc topcat} to cross-match these sources with the TIC v8.2 and to select 629 stars within the selection boundary, brighter than $V = 13$ mag, and not previously selected from any of the catalogues described above. We used visual inspection of the TESS light curves for these 629 stars to identify and measure the orbital periods of 126 detached eclipsing binary stars with variability between the eclipses due to effects other than pulsations with an amplitude less than 1\%. Of these 127 DEBs, 2 are P1 stars. 

The four transiting brown dwarfs selected for inclusion  in {\tt scv1a} are a subset of the systems selected for subsample {\tt scv6}, described below in section \ref{sec:scv6}.

\subsubsection{{\tt scv1b}: Astrometric binary stars}

The {\tt scv1b} subsample consists of astrometric systems from the Gaia Mission (DR3). \cite{Halbwachs2023} and \cite{Holl2023} reported the inclinations of these systems from a combined orbital solution, determined from an SB1 spectroscopic orbit in combination with astrometry \citep[see also][for the Non Single Star, NSS, catalog in Gaia DR3]{Gaia2023b}. 
Systems with a known inclination are important calibrators for asteroseismology and stellar astrophysics. Because of the small number of eclipsing binary systems that host oscillators, this sample provides a large set of alternative calibrator objects. Some of these systems may also be very long-period eclipsing binaries, e.g. this sample includes P1 stars in an eclipsing binary with a 65-day orbit showing solar-like oscillations (Gaia DR3 4785405080341786368) from \cite{Beck2024}.
The feasibility of the {\tt scv1b} strategy was recently demonstrated by \citet{Beck2026}, who used an astrometric double-lined binary as a proof of concept for testing asteroseismic masses.

There are 1539 stars within the selection boundary of LOPS2. Systems with orbital inclinations close to $90^{\circ}$ (i.e., an edge-on orientation) are preferred because the mass estimates will then be more precise. 

\subsubsection{{\tt scv1c}: Wide binary stars}

We identified a sample of wide, co-moving binaries in common between the independent catalogues of \cite{hartman_lepine2020} and \cite{El-Badry2021}. Stars included in one of the PLATO core-science samples as listed in the PLATO input catalogue version 2.1 (PIC2.1) were then selected for inclusion in subsample {\tt scv1c}.

These wide binary stars can be used for the validation of PLATO age estimates under the key assumption that the components are coeval. An example of the application of this technique for two very bright {\it Kepler} binaries can be found in \cite{SilvaAguirre2017}. Systems made up of a bright FGK star, i.e. a star in the P1 or P2 core science samples,  and an M-dwarf companion, i.e. a P4 star, are particularly valuable because the seismic age of the primary constrains fairly tightly that of the M dwarf. 

Furthermore, such binaries are increasingly used by spectroscopic surveys to validate their chemical abundance estimates \citep[e.g.][]{meszaros2025}. Therefore, {\tt scv1c} offers an opportunity to assess the precision of the metallicites inferred for the PLATO targets \citep{Gent2022,olander2025} that are an essential ingredient of any seismic inferences \citep[e.g.][]{Chaplin2014}. 



Finally, the binary systems that are unresolved by the PLATO cameras
 will also be valuable for understanding the impact of unresolved or marginally-resolved sources on the processing and interpretation of the mission data products \citep{white2017}. 

\subsubsection{{\tt scv1d}: HW Vir-type binaries}
HW Vir is a hot subdwarf star that shows deep, sharp eclipses from a low-mass companion every 2.8 hours. The time of mid-eclipse can be measured with a precision of about 1 second using ground-based observations of this post-common envelope eclipsing binary. Observations of HW Vir from the ground and space over two observing seasons were used to validate the accuracy of the time-stamps on data from the CHEOPS mission \citep{Fortier2024}. HW Vir is not in the LOPS2 field but other post-common envelope eclipsing binaries can be used to validate the absolute time accuracy of PLATO data to better than one second.

Stars have been selected from \cite{Schaffenroth2022}. Only AA Dor (V=11.15) from this catalogue lies within the LOPS2 selection boundary and is bright enough to satisfy the magnitude requirement (i.e., $6 \leq m_V \leq 13$ mag). 
From eclipse timing measurements obtained with the TESS mission, \cite{Baran2021} were able to confirm that the orbital period is stable, with an upper limit to any period change of $5.75 \cdot 10^{-13}s \, s^{-1}$, so only a few ground-based observations will be needed to update the eclipse ephemeris and test the accuracy of the PLATO time stamps. 

\subsubsection{{\tt scv1e}: Wide white dwarf binaries}
White dwarf companions to PLATO P1 and P2 targets on wide orbits, i.e. systems where there has not been any mass transfer between the stars, can be used to test for consistency between the age of a star estimated using asteroseismology and the age of its white dwarf companion estimated from the time for it to cool to its current effective temperature plus an estimate of its main-sequence lifetime \citep{Rebassa_Mansergas_2021}.
Targets for subsample {\tt scv1e} have been selected from the catalogue of white-dwarf – main-sequence star binaries identified using Gaia DR3 data published by \cite{El-Badry2021}. We cross-matched the main-sequence stars in this catalogue with the P1 sample and retained only those stars with a probability $<10$\% to be a chance alignment with the nearby white dwarf. The white dwarf stars are much too faint for useful data to be obtained using PLATO so only the main-sequence P1 stars are included in this subsample.


\subsection{Legacy and benchmark stars ({\tt scv2})}

As part of the PLATO effort, the Plato Benchmark Stars working group \citep{Merle2026} have consolidated a list of so-called benchmark-stars to help prepare, validate and/or calibrate PLATO pipelines that will be reun at the PLATO Data Center (PDC).  While many of these targets do not need to be observed by PLATO for the preparation and testing of the stellar pipelines, a subset of these will necessarily be observed by PLATO in order to validate the stellar pipelines.  The {\tt scv2} target list has been compiled from this latter subset, and comprises two subsamples: {\tt scv2a} includes targets with precise interferometric radii and {\tt scv2b} comprises legacy and benchmark stars, such as $\alpha\,Cen\,A$.
The {\tt scv2b} list contains solar-twins and solar-like stars, standard or single bright stars, stars with sun-like oscillations, visual binaries, members of open clusters, the so-called Titan stars \citep{Giribaldi2021} and stars from the Gaia Data Release 3 Golden Sample \citep{Gaia-Creevey2023}.  These are all discussed further below.

\subsubsection{Sample Source Description}\label{sec:scv2-selection}

For the selection of targets, we describe below our approach to compile literature data for the PLATO Benchmark Stars, i.e. targets that are not limited to the PLATO field of view. In Sect.~\ref{sec:svc2-char} we discuss the characteristics of the {\tt scv2} sample, i.e. those stars within LOPS2.   

For the ``interferometric'' targets in {\tt scv2a}, we searched the literature and selected stars which have a measured angular diameter.   While we focused on finding sources with both interferometric and seismic data, where possible, the selection was not limited to this.  Our selection includes the Gaia FGK Benchmark stars \citep{Soubiran2024} and targets available in \cite{Bazot2011,Huber2012, White2013,Kervella2017,Bond2017,Bond2020} and \cite{Gardner2018}.
A set of M dwarfs with interferometric data was also included from \cite{Boyajian2012}, but we have updated these parameters by exploiting the astrophysical parameters from the latest Gaia Data Release \citep{Creevey2023}.
We also cross-matched the PIC (version 2.2.0.1) for LOPS2 with Gaia DR3 to find targets specifically in the PLATO field of view that are measurable using interferometry, using both the VLTI and the CHARA array.  This comprises a total of 50 stars. 

For the selection of the {\tt scv2b} stars we describe them by star type.  
Solar twins are stars considered to have spectra identical to the Sun’s, while solar-type stars are considered those with stellar parameters similar to that of the Sun, e.g. radius, mass, $T_{\rm eff}$, and luminosity within 5\% – 10\% of the Sun’s parameters. These sources were collected from a number of catalogues, see details in the PLATO Benchmark Database \citep{Merle2026}. Examples include 27 stars selected from the Gaia Golden sample of astrophysical parameters solar-like stars (see below), \cite{Melendez2012} and \citep{Graczyk2016}.

Standard or single bright stars have been well measured for many years with many different instruments, thus providing true benchmark tests for stellar physics. Most of these stars are part of the Gaia FGK benchmark stars \citep{Soubiran2024}. Additional sources for these stars come from \cite{Maxted2020,Maxted2024}.

Seismic stars are those with either measured asteroseismic frequencies or at minimum the mean seismic quantities ($\nu_{\rm max}, \, \langle\Delta\nu\rangle$). These stars were compiled from many catalogues of seismic data, observed mainly with {\it Kepler}, TESS, CoRoT, and also ground-based telescopes. The sample contains primarily main sequence and subgiant stars. The objects were selected from the following publications: \cite{Mathur2012,Creevey2017,Lund2017,SilvaAguirre2017,Li2020,Zhou2024}. For \cite{Zhou2024}, the astrophysical parameters were taken from \cite{Gaia-Creevey2023}.


Visual binaries have component stars that are well separated on the sky such that we can consider them as single stars for interpreting their observations. The interest in these is the extra constraints provided by the orbital system parameters. Example references for these are \cite{Valenti2005}, \cite{Kervella2017}. These comprise only very few of the sample.

The Titan stars refer to mildly metal-poor stars from the catalogues of \cite{Giribaldi2021} and \cite{Giribaldi2023}. Their metallicities span a range of -0.75 to -0.30 dex.   Their \teff\ have been measured by analysing the H$\alpha$ line, and they have been carefully compared with other methods to provide truly accurate \teff.

Based on a list of over 43,000 members of open clusters in LOPS2 provided within the consortium (L. Briganti private comm.), we selected a representative sample of main sequence and subgiant targets manually by choosing the brightest targets covering the span in luminosity and effective temperatures of the clusters. The result is a list of clusters with members of between 4 and 30 stars, to give a total sample of 193 targets.

The Golden sample of Gaia astrophysical parameters \citep{Gaia-Creevey2023} is a catalogue of very high precision parameters available in the Gaia archive (DR3). The members of the catalogue are selected based on high quality parallactic, photometric, and parameter information. It contains around four million stars spanning the main sequence, sub-giant, and giant evolutionary phases. Among this catalogue, we identified FGK dwarfs and subgiants with $G < 9.1$ mag in LOPS2, and separated the stars in bins of $T_{\rm eff}$ and log $g$ in order to identify the brightest targets in each bin. The selected FGK dwarfs comprise about 60 stars. The M dwarfs in the Gaia golden sample are much fainter, and so we selected about 40 stars with $13 < G < 16$.
In addition to these, 27 solar-like stars were selected from the table \texttt{gaiadr3.gold\_sample\_solar\_analogue}, which contains $\sim$5000 solar-like candidates.  The 27 selected stars have the closest parameters to the Sun, have published RVS spectra, and are the brightest in the sample.


\subsubsection{Sample characteristics}\label{sec:svc2-char}

\begin{figure}
    \centering
    
\includegraphics[width=0.48\linewidth]{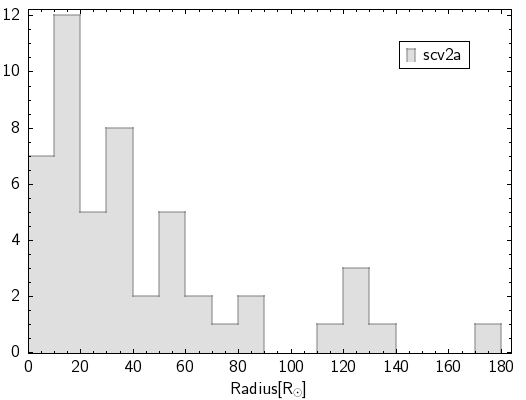}
\includegraphics[width=0.48\linewidth]{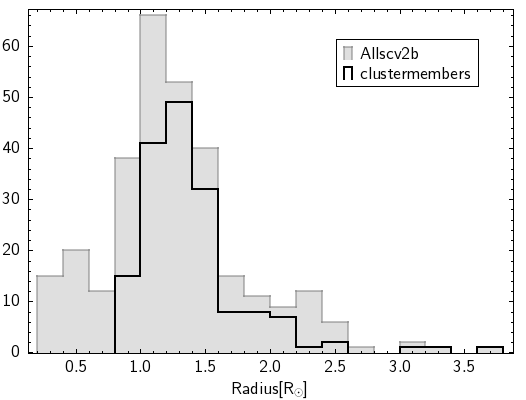}
\includegraphics[width=0.48\linewidth]{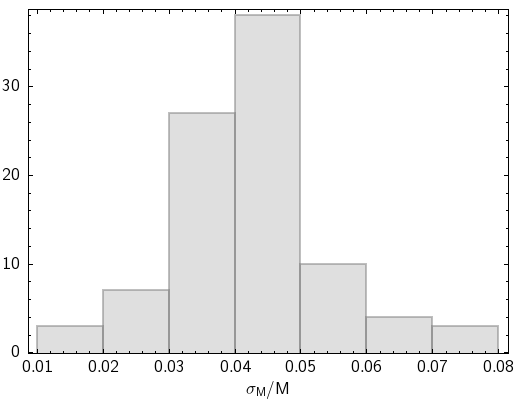}
\includegraphics[width=0.48\linewidth]{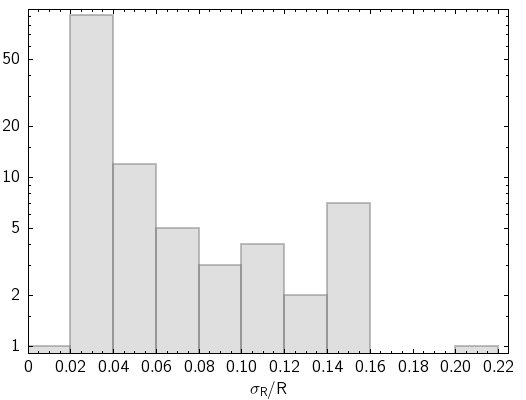}
    \caption{Top panels: Histograms of radius properties of the {\tt scv2a} (left) and {\tt scv2b} (right) sample. On the right panel we also overplot the subsample of the {\tt scv2b} targets that are members of clusters.
    Lower panels: Relative errors in the mass and radius of the {\tt scv2b} sample as characterised by \cite{Merle2026}.}
    \label{fig:scv2_hist}
\end{figure}

\begin{figure}
    \centering
    \includegraphics[width=0.48\linewidth]{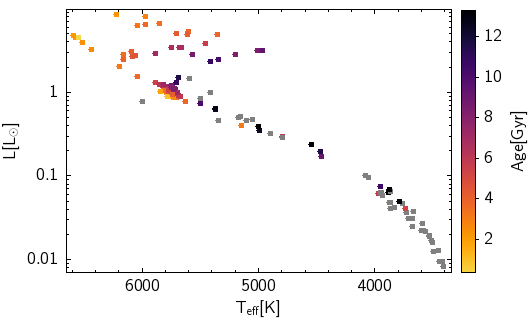}
    \includegraphics[width=0.48\linewidth]{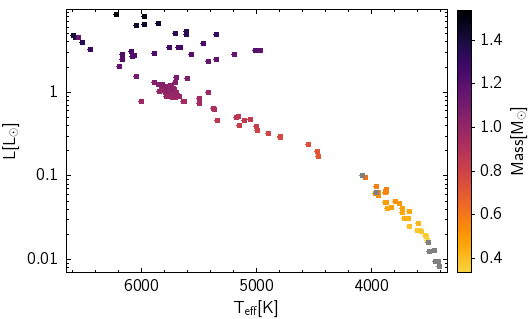}
    \includegraphics[width=0.48\linewidth]{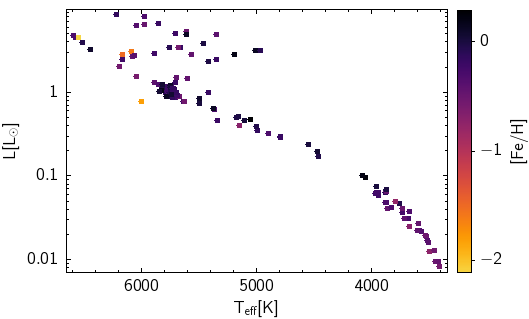}
    \includegraphics[width=0.48\linewidth]{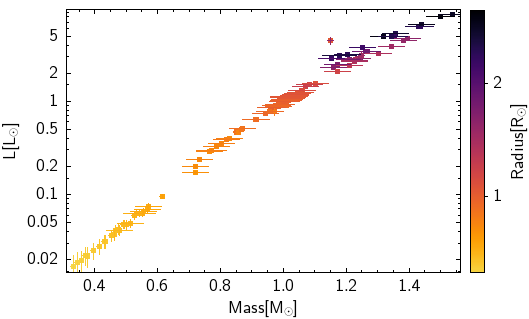}
    \caption{From top left to bottom right:  HR diagram showing the distribution of the {\tt scv2b} targets, colour-coded by age (top left), mass (top right) and metallicity (lower left).  Stellar properties are from \cite{Merle2026}, and grey symbols indicate that no property has yet been attributed to that star.  Lower right: Distribution of {\tt scv2b} targets highlighting  the mass-luminosity relation, according to the properties from \cite{Merle2026}.}
    \label{fig:scv2b}
\end{figure}

We illustrate the final selected sample properties of the {\tt scv2a} and {\tt scv2b} targets in Figures \ref{fig:scv2_hist} and \ref{fig:scv2b}.
In Fig.~\ref{fig:scv2_hist} we show the distribution of the radius of the two samples.
As the LOPS2 field is in the south, it is primarily the VLTI interferometer that can observe the {\tt scv2a} (interferometric) targets. The VLTI has a shorter baseline compared to CHARA in the northern hemisphere, and therefore the observable targets have larger apparent diameters.  This naturally restricts the observations primarily to larger stars, i.e. subgiants and giants.  The radii of these stars spans 0.5 -- 175 $R_{\odot}$ according to the characterisation of the PIC catalogue, as can be seen from the top left panel.
The {\tt scv2b} targets are primarily a sample of main sequence and subgiant stars, and the distribution of their radii are shown in the top right panel of this same figure.  
The radii spans 0.2 -- 3.6 $R_{\odot}$ according to the PIC properties.   The inset shows the
histogram of the sample of the {\tt scv2b} stars that are members of clusters.  
A total of 171 stars are members of 36 clusters in the PLATO field of view.

In the lower panel of Fig.~\ref{fig:scv2_hist} we show the currently known relative uncertainty on the mass and radius properties of the {\tt scv2b} targets.
These stars are the targets of current scientific studies with the aim to reach less than 3\% uncertainties on the radius.

In Fig.~\ref{fig:scv2b} we illustrate the physical properties of the {\tt scv2b} sample in the form of HR diagrams and the mass-luminosity relation. 
HR diagrams are illustrated with colour-codes according to other stellar properties: top left and right panels indicate the age and mass as the colour-code, where these stellar properties were taken from \cite{Merle2026}.  The lower left shows the HR diagram with metallicity as the colour-code and these values for this figure were obtained using the Gaia DR3 mh\_gspphot value.  Grey points are stars where there is no corresponding stellar property  yet in \cite{Merle2026}.

Finally, in the lower right panel we show the mass-luminosity relation for the {\tt scv2b} sample, including also the known error bars on both of these values.  Again these properties have been taken from \cite{Merle2026}.

\subsection{Photometrically stable stars ({\tt scv3})}

For the long-term calibration of PLATO photometry, photometrically stable stars are required. Short- and long-period rotationally variable magnetic chemically peculiar (Ap/Bp) stars were identified as suitable targets for that purpose.

The basic idea behind this selection is to take advantage of the fact that Ap/Bp stars possess an inhomogeneous, but highly stable surface spot distribution. The stability of surface structures on these stars is linked to the constant fossil magnetic fields present in their interiors and permeating their atmospheres. 
According to observational studies, conducted for some of Ap/Bp stars over many decades \citep[e.g.][]{Krticka2017}, and theoretical modelling \citep{Braithwaite2004,Duez2010}, one confidently expects the stability of their magnetic fields and related surface spots over time scales of several years, which will be covered by PLATO’s long-duration observations. 
These stars will therefore exhibit a very stable photometric behaviour when observed by PLATO – a simple (typically single or double-wave), mono-periodic rotational variation with $P_{\rm rot}$ of the order of few days without any changes in either phase or amplitude. It would be straightforward to remove this variability from the observed PLATO light curves, with any residual variation highlighting uncorrected instrumental artefacts and long-term drifts.

Moreover, a small subset of Ap/Bp stars is known to have very long rotational periods \citep{Mathys2017,Mathys2020,Huemmerich2024}, reaching $\sim$10–100 yr in exceptional cases. Such super-slowly rotating Ap/Bp (ssrAp) stars will appear as truly constant stars (for $P_{\rm rot} \gg 2$ yr) or as targets showing only slow trends (for $P_{\rm rot} \gtrsim 2$ yr) of brightness during the few years of PLATO monitoring of the long observation pointing fields (LOPs).
It will be possible to use this group of Ap/Bp stars as photometric standards directly, without pre-whitening their rotational variability.

Independently of the value of rotational period, Ap/Bp stars are easy to identify spectroscopically by the
presence of anomalously strong or (in some cases) weak lines of certain chemical elements (e.g. Sr, Cr, Eu for
A-type stars, He and Si for B-type stars) in their spectra. These spectral anomalies, which are straightforward
to recognise in low-resolution classification spectra, are the basis for the spectral classification of these stars as ``Ap/Bp" in the literature. The success rate of detecting magnetic fields in chemically peculiar stars based on their spectral peculiarity types is known to be nearly 100\% \citep[e.g.][]{Thomson-Paressant2024}, allowing us to rely on spectral classification catalogues for selecting targets in the PLATO LOP fields without the need of direct magnetic field observations.

\subsubsection{{\tt scv3} sample selection}
The targets selected as long-term photometric calibrators belong to one of two groups satisfying the following conditions. (i) Targets in {\tt scv3a} are photometrically constant stars in the visual magnitude range from $6 \leq m_V \leq 13$, that are dwarfs of spectral type A and B, including Ap/Bp stars. They are photometrically constant at a level of 50 ppm/hr or lower and do not have any known or measurable variability on timescales of up to two years. (ii) Targets in {\tt scv3b} are Ap/Bp stars with stable periodic variability in the visual magnitude range from $6 \leq m_V \leq 13$, that are dwarfs of spectral type Bp or Ap. They exhibit stable mono-periodic rotational modulation. For each of the two groups, {\tt scv3a} and {\tt scv3b}, a maximum of 100 targets per LOP field will be observed. 

While there are chemically normal B and A stars without any pulsational or rotational variability \citep[e.g.][]{Murphy2015}, finding these objects among the plethora of variable early-type stars requires space photometry data sets comparable in quality and time span to the existing {\it Kepler} or future PLATO LOPs observations. In the absence of such data for LOPS2, we must resort to using only well-established or suspected ssrAp stars for {\tt scv3a}. The second group of calibration targets, {\tt scv3b}, by definition, contains only short-period Ap/Bp stars.

We have used the stellar spectral classifications from \cite{Renson2009} and \cite{Skiff2014}, augmented with several recent TESS Ap/Bp-star survey articles \citep{Cunha2019,Holdsworth2021,Holdsworth2024}, to create a list of targets according to the requirements outlined above. In terms of spectral classification, we required the stellar spectral type to fall within the B–F range (cool Ap stars extend to early-F spectral types) and to include at least one of the following anomalous elements: Si, Sr, Cr, Eu, or He. A small number of late-B HgMn stars, typically characterised by one of these elements combined in the spectral designation with Hg or/and Mn, was identified and excluded. The latter type of chemically peculiar stars is known to lack strong magnetic fields \citep{Makaganiuk2011} and exhibit slowly evolving surface abundance spots \citep{Kochukhov2007}, resulting in changing rotational phase curves \citep{Niemczura}. These characteristics make HgMn stars unsuitable for the purpose of long-term photometric calibration.

The merged catalogue of spectroscopically identified Ap/Bp stars, constructed from the sources listed above, is limited to relatively bright stars. Specifically, it contains very few objects in the 11–13 mag range. To overcome this limitation, we investigated the possibility of complementing these bright targets with fainter Ap/Bp stars identified by \cite{Paunzen2022} using Gaia BP/RP spectrophotometry. However, since their methodology of finding Ap/Bp stars is relatively new and has not been fully validated, we process the candidate list from \cite{Paunzen2022} separately from the spectroscopically confirmed Ap/Bp stars.

\subsubsection{{\tt scv3} targets in LOPS2}
In LOPS2 we identified 336 spectroscopically confirmed Ap/Bp targets with a mean Gaia magnitude $G = 9.1$ and 512 candidate Ap/Bp stars using Gaia spectrophotometry with a mean magnitude $G = 12$. We cross-matched all these targets with the Gaia DR3 data \citep{Gaia2022-Vizier}, retrieving Gaia magnitudes, colours, coordinates, RUWE, and DR3 identifications. 
Finally, Ap/Bp targets in LOPS2 were cross-matched against the lists of established and candidate ssrAp stars \citep{Mathys2020,Mathys2022,Mathys2024,Huemmerich2024}, resulting in 5 established and 17 candidate long-period Ap/Bp stars within the $6 \leq m_V \leq 13$ magnitude range.

We expect some fraction of the candidate {\tt scv3} targets in LOPS2 to have bright nearby companions that would contaminate the Ap/Bp star light curve to some extent, potentially making it problematic to use as a stable reference. Given the abundance of suitable targets in LOPS2, we adopt a conservative approach of removing all significantly contaminated targets from the {\tt scv3} list rather than investigating the nature of contaminants in each case.

Approximate contamination factors were calculated for each Ap/Bp target based on the Gaia data. To this end, we queried the Gaia DR3 catalogue for the list of sources around each target of interest. Gaia $G$ magnitudes were converted to PLATO $P$ magnitudes. Then, each target was modelled with a Gaussian point spread function assuming 80\% of the flux to be enclosed within the $2 \times 2$ pixel window. The contamination factor was defined as the ratio of the main target flux in the $4 \times 4$ pixel window centred on the target to the sum of fluxes of all contaminants in this window. Setting the contamination threshold to $< 0.1$ (i.e. the main target contributing $>$ 90\% of the flux), we retained 285 targets from the original list of 336 bright spectroscopically confirmed Ap/Bp stars and 93 stars from the list of faint 512 Ap/Bp candidates identified using Gaia spectrophotometry.

Additionally, we examined RUWE (Renormalized Unit Weight Error) reported in the Gaia DR3 catalogue to identify potential unresolved binaries. Using RUWE $> 2$ to exclude likely complex sources, we identified further 38 targets in the bright list that are possible members of multiple systems and therefore less suitable as photometric reference targets.

\begin{figure}[t]
\centering
\includegraphics[width=0.95\textwidth]{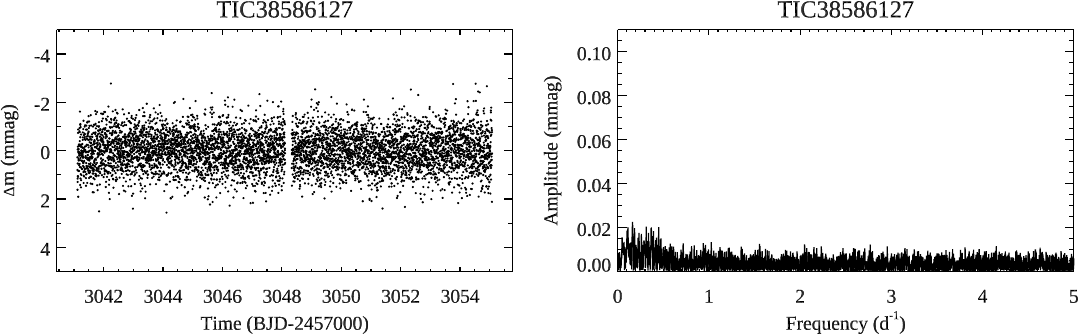}
\includegraphics[width=0.95\textwidth]{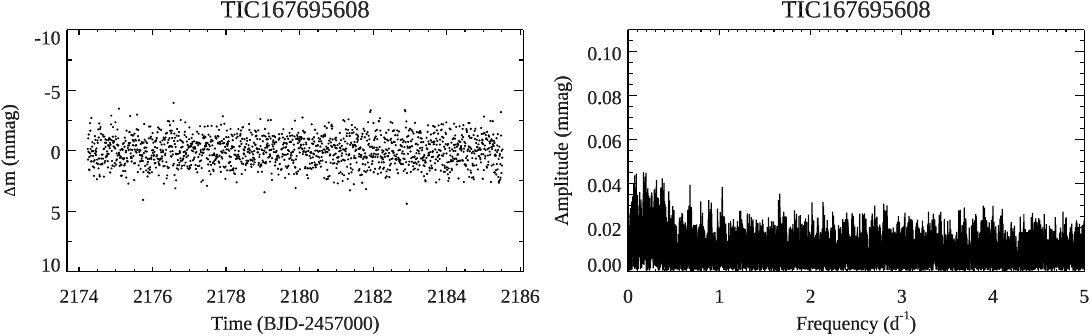}
\caption{Examples of TESS light curve analysis of {\tt scv3a} targets. 
Top panels: TESS data for one sector and amplitude spectrum for all sectors for TIC~38586127 (HD 27472). This target shows no measurable variability in TESS data and is therefore included in the {\tt scv3a} sample.
Bottom panels: the same for the fainter {\tt scv3a} star TIC~167695608 (TYC 8912-1407-1).}
\label{fig:scv3_1}
\end{figure}

\subsubsection{TESS light curves of {\tt scv3} targets}
We analysed all 336 bright spectroscopic Ap/Bp stars and 93 uncontaminated faint Ap/Bp candidates using TESS data. The goal of this analysis was to identify rotational modulation, measure rotational periods, and examine residual signals. This allows us to identify close binaries with variable components as well as stars showing complex variability (e.g. a combination of rotational modulation and pulsations), making them unsuitable to serve as reference photometric targets.

For each star, all available TESS light curves (up to sector 69) were downloaded and processed with custom time-series algorithms \citep{Kochukhov2021}. This analysis included fitting a mono-periodic harmonic model to the stellar rotational signal in all TESS sectors simultaneously, while fitting the background in each sector with a low-degree polynomial. The lowest frequency that could be recovered with this procedure is approximately equal to $0.04 \,d^{-1}$, corresponding to the length of one TESS sector.

The following criteria were used to judge suitability of targets for inclusion in the {\tt scv3a} and {\tt scv3b} samples:
\begin{itemize}
\item {\tt scv3a} candidate: No rotational modulation visible. Any signals in the amplitude spectrum in the [0.04, 20]
$d^{-1}$ frequency range are below 0.5 mmag.
\item {\tt scv3b} candidate: Rotational modulation clearly present. The residual amplitude peaks, after pre-whitening
the rotational modulation, are below 0.5 mmag in the [0.04, 20] $d^{-1}$ frequency range.
\end{itemize}

Whether or not these stars will meet the 49ppm/hr stability requirement could be evaluated only having the PLATO light curves collected over the entire LOP duration. The $\leq \, 20$~$d^{-1}$ frequency limit is adopted here to exclude known types of stellar variability (rotation, $\delta$ Scuti, SPB, $\beta$ Cep pulsations) that could be encountered in the target stars themselves or in their contaminants. The only other known type of pulsational variability outside this frequency range – rapidly oscillating Ap stars (50–300 $d^{-1}$) – occurs at much higher frequencies and is easy to average out.

\begin{figure}[!h]
\centering
\includegraphics[width=0.9\textwidth]{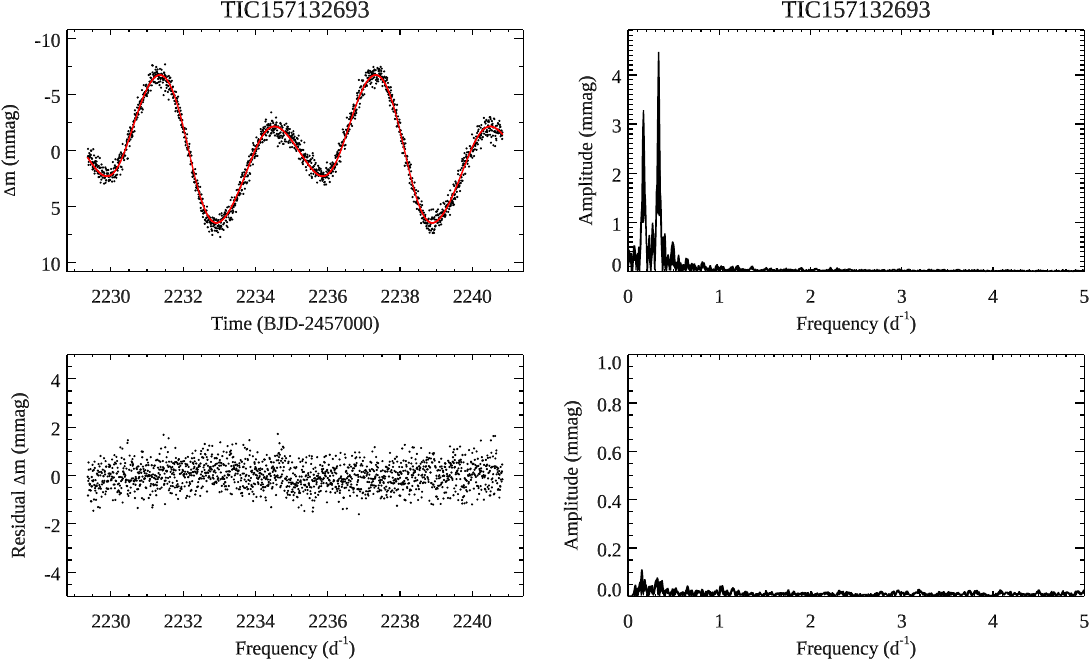}
\caption{
TESS data for one sector and amplitude spectrum for all sectors for TIC~157132693 (CD-40 2835). Upper panels show original light curve, bottom panels correspond to the analysis of residual data after fitting and removing the rotational modulation. TIC~157132693 is judged suitable for {\tt scv3b} given the absence of variability in the residual light curve. }
\label{fig:scv3_2a}
\end{figure}

\begin{figure}[!h]
\centering
\includegraphics[width=0.9\textwidth]{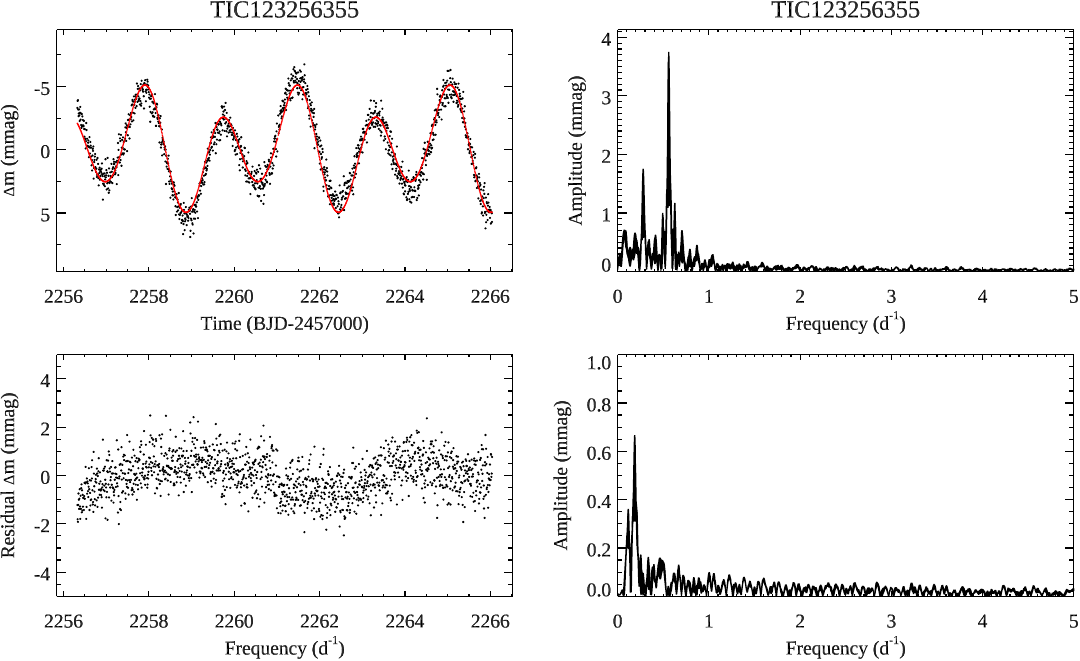}
\caption{
Same as Figure \ref{fig:scv3_2a} but for TIC~123256355 (CD$-$46 3282). In this case, the residual light curve reveals coherent low-frequency signal. This target is therefore judged as contaminated and unsuitable for inclusion in the {\tt scv3b} sample.}
\label{fig:scv3_2b}
\end{figure}

The full set of results from this time-series analysis will be reported elsewhere. Some examples of TESS light curves for the {\tt scv3a} and {\tt scv3b} candidates are illustrated in Figs. \ref{fig:scv3_1}, \ref{fig:scv3_2a} and \ref{fig:scv3_2b}.

The outcome of the time-series analysis of the TESS data of Ap/Bp stars in LOPS2 is the following:
\begin{itemize}
\item 54 targets among spectroscopic Ap/Bp stars satisfy the criteria for inclusion in the {\tt scv3a} sample,
\item 173 targets among spectroscopic Ap/Bp stars satisfy the criteria for inclusion in the {\tt scv3b} sample,
\item 18 targets among Gaia Ap/Bp candidates satisfy the criteria for inclusion in the {\tt scv3b} sample. However,
many targets from this list exhibited photometric variability not typical of Ap/Bp stars, suggesting that the
selection of Ap/Bp stars based on Gaia spectrophotometry is not reliable. In the light of these results, we
do not consider photometric Ap/Bp candidates in subsequent analysis.
\end{itemize}

This assessment was carried out under the assumption that the photometric behaviour of LOPS2 targets in TESS data, in particular the degree of mutual contamination of nearby targets, represents a good proxy of the expected PLATO signals. This is a reasonable assumption considering similar pixel scale (15" for PLATO and 21" for TESS) of the two missions. On the other hand, the PLATO PSF will extend over a smaller number of pixels. Moreover, PLATO’s band pass will be shifted to bluer wavelengths relative to that of TESS, which will increase the amplitude of typical rotational variability of Ap/Bp stars. These factors will reduce the impact of contamination of Ap/Bp PLATO targets by random nearby stars. Therefore, examining this contamination based on TESS data represents a conservative approach.

\subsubsection{Final {\tt scv3} target selection}
Guided by the analysis presented above, we selected 54 stars for the {\tt scv3a} sample. This includes 4 previously established and 12 candidate ssrAp stars.

We found that 173 short-period Ap/Bp stars in LOPS2 are suitable for the {\tt scv3b} sample; 100 of these objects are selected as prime targets (priority 1), with the remaining 73 (priority 2) kept in reserve. In this selection, we attempted to achieve an approximately uniform distribution of {\tt scv3b} targets across the field of view and assigned lower priority to targets close to the edge of LOPS2. No other considerations were taken into account.

\begin{figure}[t]
\centering
\includegraphics[width=6.4cm]{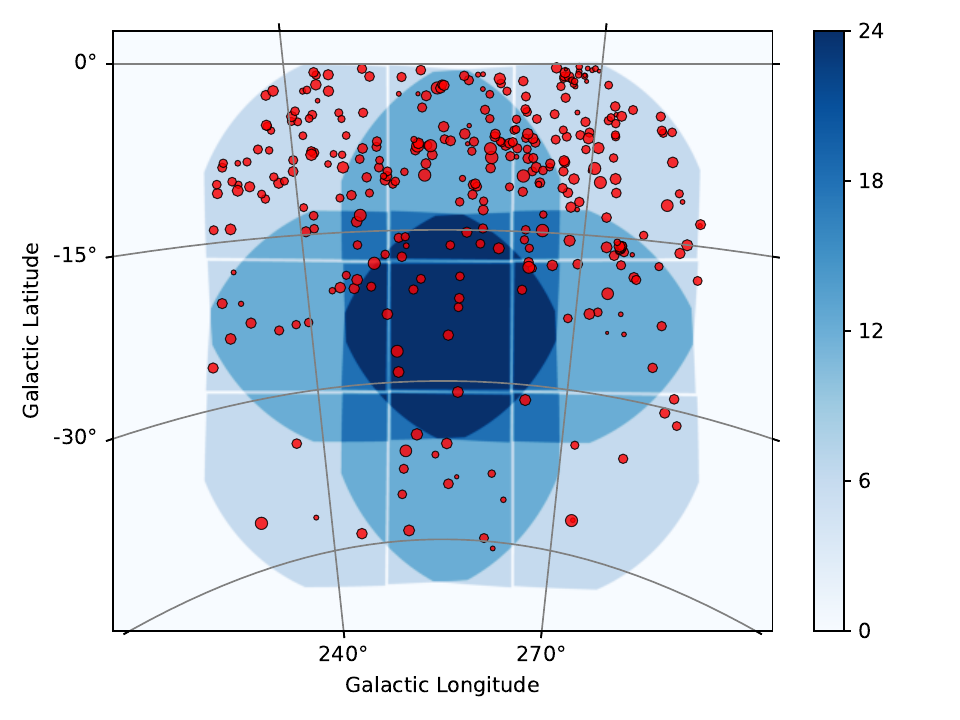}
\includegraphics[width=6.4cm]{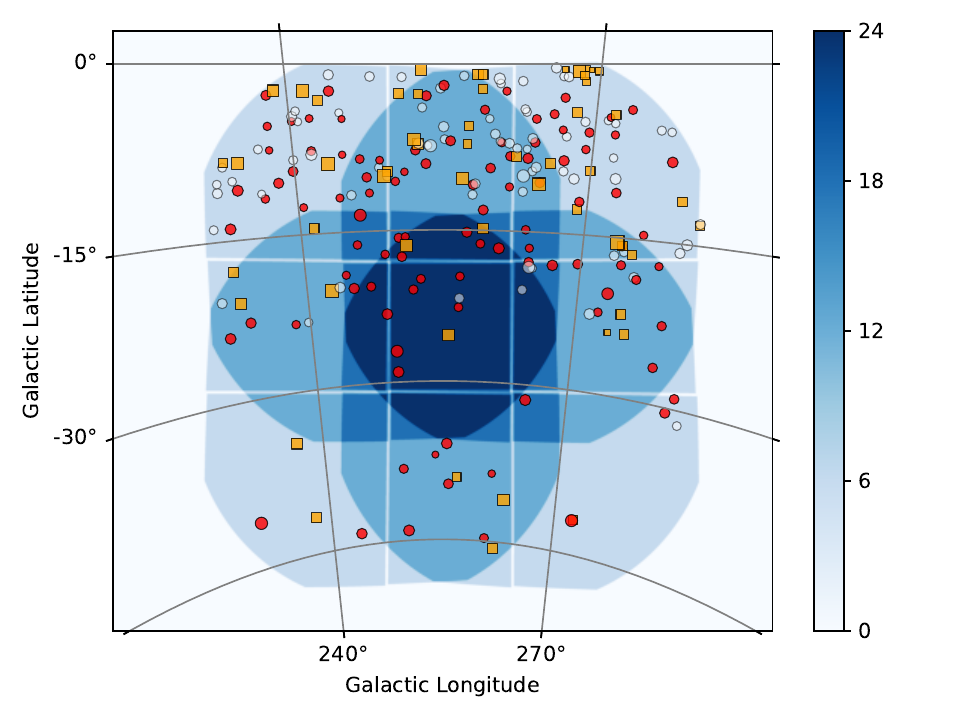}
\caption{Distribution of the {\tt scv3} targets in LOPS2. Left panel: initial sample of the spectroscopically confirmed Ap/Bp stars. Right panel: the final selection of scv3a (squares), scv3b first priority (filled circles) and second priority (open circles) targets. In both panels the size of the symbols is proportional to target brightness and the different shades of blue (see colourbars) show the areas monitored by 6, 12, 18 and 24 normal cameras (N-CAMs), respectively.}
\label{fig:scv3_3}
\end{figure}

Distribution of the initial sample of spectroscopic Ap/Bp stars in the LOPS2 field and our final selection of the {\tt scv3a} and {\tt scv3b} targets is shown in Fig.~\ref{fig:scv3_3}.

\subsection{Red Giants ({\tt scv4})}
\newcommand{\modif}[1]{\textcolor{blue}{#1}}

\subsubsection{Seismic constraints from ageing stars}

Evolved stars are an essential part of PLATO's science calibration activities, since their seismic study allows us to test physical phenomena in their main sequence (MS) progenitor stars. 
This unique information comes from the oscillations that probe the inner stellar regions. However, gravity waves that propagate in the radiative core of the Sun or of MS stars remain hidden since they cannot couple efficiently with pressure waves \citep{2018A&A...617A.108A}. 
As a result, solar-like oscillations in MS stars are quite insensitive to physical conditions in the core region. In contrast, mode coupling is efficient in evolved low-mass stars \citep{Hekker2017}. The resulting modes, called mixed modes, reveal the physical properties in the radiative core, but also reveal past conditions along stellar evolution. 

As predicted by \cite{2013ApJ...766..118M}, the period spacing of gravity ($g$-) modes during the core-Helium burning (CHeB) phase allows us to characterize the convection properties during MS. 
With \textsl{Kepler} observations, \cite{2015MNRAS.453.2290B,2017MNRAS.469.4718B} could test different scenarios of convective mixing in those CHeB stars. They showed a slight mass and a significant metallicity dependence, which must be addressed by PLATO observations to constrain stellar evolution models in large mass and metallicity ranges and ensure unbiased ages during the MS phase. Observations revealed complex signatures in the mixed-mode pattern that can be used to refine the physics of core convection  
\citep{2022NatCo..13.7553V}. Similarly, the initial helium and the helium core mass can be estimated in the secondary red clump, and then mapped against that of the MS stars. 
Therefore, {\tt scv4} calibration stars are mostly CHeB stars with different masses and metallicities. The {\tt scv4} sample is defined to provide the most stringent constraints from ageing stars.

\subsubsection{{\tt scv4} sample selection criteria}


The survey sample is initiated with the identification of 133,106 red giants located in the LOPS2 field. These stars are derived from a cross-match between the Gaia DR3 catalog and the \texttt{StarHorse} (SH) catalog \citep{2018MNRAS.476.2556Q, 2022A&A...658A..91A}. In practice, we extracted all SH targets within LOPS2
cross-matched with the Gaia DR3 catalog. This catalog is one of the sub-surveys of the program 4MIDABLE-LR \citep{2019Msngr.175...30C}, as part of the 4MOST project  \citep{2019Msngr.175....3D}, but done at high  resolution.
For the identification of the area in the sky, we used the center coordinates of LOPS2 (see Section \ref{sec1}) and an internal circle with a ${\rm radius} = 24.324320^{\circ}$ (Nascimbeni, private communication).


The targets were selected either on the basis of their measured global seismic parameters 
\citep[$\Delta\nu$ and $\nu_{\rm max}$, see\,][for a review and definitions]{GarciaBallot2019} 
or on the high probability of detecting seismic parameters from their light curves obtained by the CoRoT, \emph{Kepler}, K2, and TESS space missions. Only stars brighter than $G = 14.5$ mag were considered. 


The appropriate calibration capacity has to cover the mass range $[0.8,3]\, M_{\odot}$ in steps of $0.1\, M_{\odot}$. These twelve mass bins ensure to cover ages within the range [0.8, 13] Gyr. In terms of metallicity, ten combinations of $Y / Z$ are needed, such that $Z \in [0.005, 0.04]$ is covered.


From simulations, we derived that 200 red giants per mass and per metallicity bin, covering the red giant branch, red clump, and secondary clump, are required for backtracking stellar evolution with rotationally split mixed modes to the main sequence phase to quantify rotation and mixing profiles. This means that the calibration process of {\tt scv4} requires $200 \times 12 \times 10 = 24\,000$ red giants to cover the full mass, age and $Y / Z$ ranges.


Among those stars, special attention was devoted to cluster members. Since PLATO offers us the possibility to gather high-precision, high-cadence, long photometric series of large samples of coeval and initially-chemically-homogeneous stars in open clusters, all red giants identified in clusters in the LOPS2 FOV were selected.

\begin{figure}[t]
\includegraphics[width=0.95\textwidth]{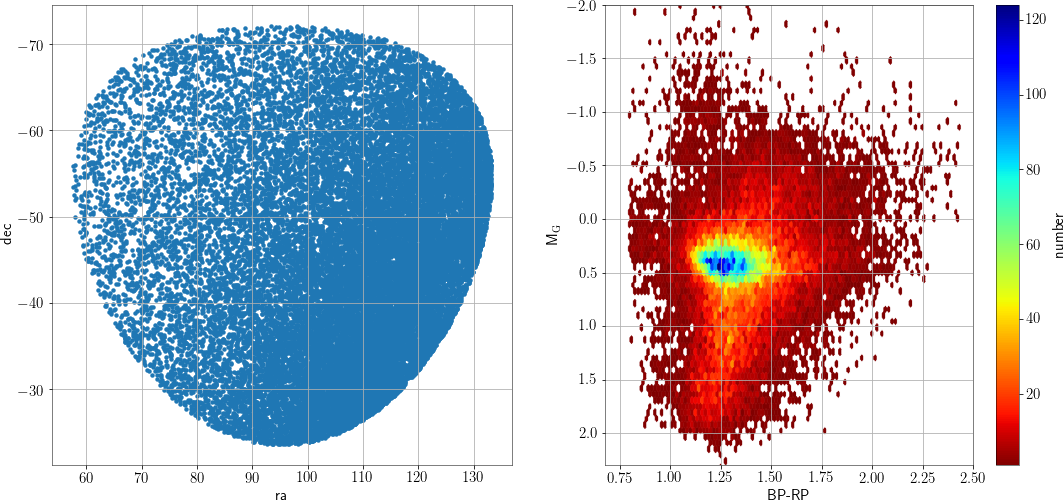}
\caption{\emph{Left:} {\tt scv4} targets in the LOPS2 field. \emph{Right:} {\tt scv4} targets in a colour-magnitude diagram}
\label{fig:LOPS2_RG}
\end{figure}

\begin{figure}[t]
\includegraphics[width=0.95\textwidth]{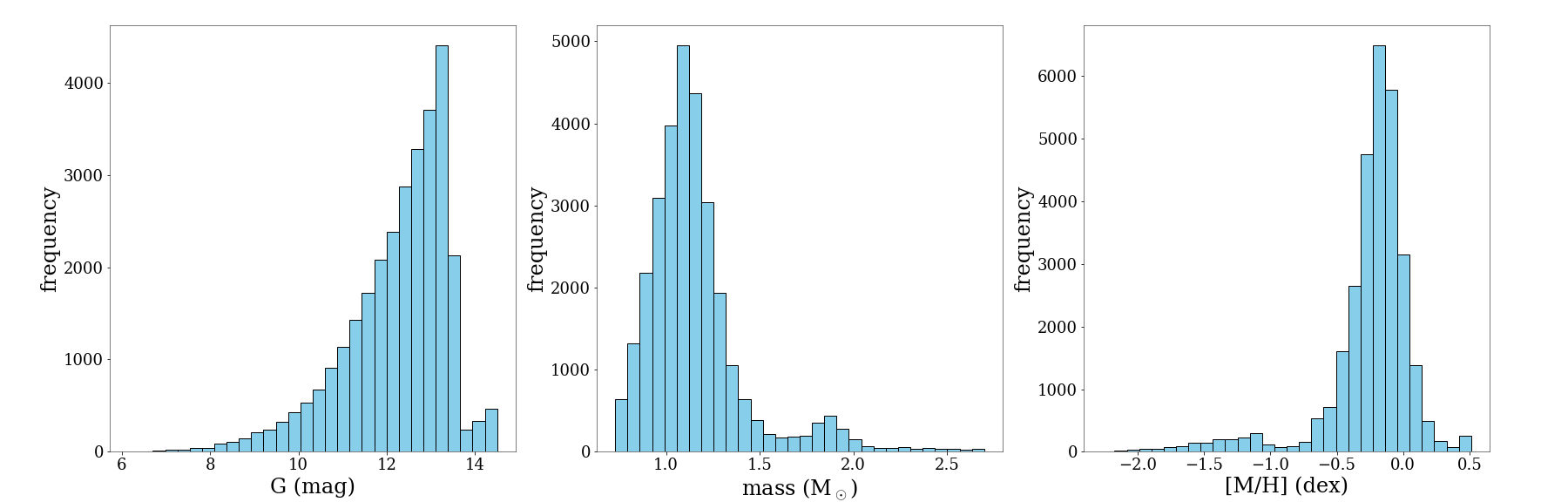}
\caption{\emph{Left:} histogram of the magnitudes of the selected targets within {\tt scv4}. \emph{Middle:} histogram of their masses. \emph{Right:} histogram of their metallicities.}
\label{fig:LOPS2_histograms}
\end{figure}

\begin{table}[t]
\caption{{\tt scv4} subsamples definition}\label{scv4_sample}%
\begin{tabular}{@{}llcc@{}}
\toprule
Subsample & Definition  & Mag. limit & Number\\
\midrule
{\tt scv4a} & RG in clusters        & 14.5 & 233  \\
{\tt scv4b} & RG with [Fe/H] $< -1$ & 14.5 & 1592  \\
{\tt scv4c} & RG with [Fe/H] $> -1$ & 13.5 & 28175  \\
\botrule
\end{tabular}
\end{table}

\subsubsection{{\tt scv4} final selection}

According to the criteria listed above, we selected all Gaia targets that fall in LOPS2 and then applied the following cuts:
\begin{itemize}
    \item $6 < G \leq 14.5$; $G_{\rm BP} - G_{\rm RP} > 0.8$;
    \item positive parallax; RUWE $<$ 1.4;
    \item $10 \leq \nu_{\textrm{max}} \leq 150\, \mu\textrm{Hz}$ (from SH and scaling relations).
\end{itemize}

In order to guarantee a sufficient coverage of a large metallicity range, we kept in the low-metallicity sample (i.e., with [M/H]$< -1.0$) all the red giant stars at $G \,<\, 14.5$ mag with \texttt{StarHorse} [M/H] $< -1$ dex. 
For [M/H] $> -1.0$ we kept only targets brighter than G $< 13.5$ mag to achieve the required number of calibration stars.
From this, we define three sub-samples within {\tt scv4}, as summarized in Table \ref{scv4_sample}, all stars identified in open clusters being kept. 

All stars in the subsamples {\tt scv4a} and {\tt scv4b} were assigned priority 1 (highest priority), as well as the first 24\,000 stars of {\tt scv4c}. The dimmest remaining stars of {\tt scv4c} were assigned priority 2.
The selected targets are shown in Fig.~{\ref{fig:LOPS2_RG}};  the histograms of their properties are shown in {Fig.~\ref{fig:LOPS2_histograms}}.


\subsection{$\gamma$ Doradus stars ({\tt scv5})}

The purpose of including $\gamma$ Doradus pulsators in the {\tt scvPIC} is to provide calibrations of stellar structure and evolution (SSE) models on the front of internal rotation and mixing profiles. This becomes possible thanks to asteroseismic modelling of gravity modes in $\gamma$ Doradus stars, a method that has been developed independently by \cite{VanReeth2016} and \cite{Ouazzani2017}. Forward asteroseismic modelling of such pulsators is particularly powerful to probe the physical circumstances in the region between the convective core and radiative envelope, where the high-order gravity modes have their dominant probing power \citep{Mombarg2019,Mombarg2020,Mombarg2021}.
Application of this method allows one to calibrate internal rotation profiles $\Omega(r,t)$ inside these early F-type stars \citep{Kurtz2014,Saio2015,VanReeth2018,GangLi2019,GangLi2020} and to 
extrapolate these results for PLATO’s primary samples, which are only slightly less massive than the 
$\gamma$ Doradus stars. Extrapolation is to be understood along the stellar mass dimension, i.e. from the range of masses typical for $\gamma$ Doradus-type variables, i.e. $M \in [1.3, 1.9] M_{\odot}$, towards the regime of PLATO’s P1/P2 sample dwarfs, i.e. $M \in [1.1, 1.3] M_{\odot}$. All these stars possess a convective core during the main sequence, such that their physical circumstances adjacent to this core are expected to be similar. 
PLATO science calibration of the internal rotation measured for $\gamma$ Doradus-type pulsators requires monitoring a dedicated sample of A9 to F4 stars during the entire long pointing(s) and covering a range of metallicities, evolutionary stages, and rotation rates. These activities will deliver the currently missing data to calibrate $\Omega(r,t)$ representative of P1/P2 sample stars having convective cores
($M \geq 1.1 M_{\odot}$).

In addition to internal rotation, $\gamma$ Doradus-type pulsators play a crucial role in the calibration of internal chemical mixing profiles. Calibration of the internal mixing is harder to achieve than internal rotation because it involves a number of transport processes relying on gradients of quantities.
One of those processes in slow rotators is element transport due to atomic diffusion, which accumulates over time, i.e. $D_{\rm mix}(r,t)$, notably settling of helium \citep[He,][]{Verma2019} 
and the consequences of radiative levitation active in the hotter P1/P2 sample stars. Ignoring microscopic mixing due to atomic diffusion may lead to mass, radius, and age uncertainties of $\sim$5\%, $\sim$2\%, and $\sim$25\%, respectively, for P1/P2 dwarfs with $M \geq 1.2 M_{\odot}$ \citep{Deal2020}. These authors showed that rotational mixing and radiative levitation act together and have opposite effects for the P1/P2 stars with a convective core. 
Along with rotational mixing, this may have a major effect on the stellar age-dating \citep{Mombarg2022,Mombarg2024a,Mombarg2024b}. It is therefore needed to calibrate the levels of mixing and improve the stellar models on this front, particularly near the convective core of P1/P2 stars with $M \geq 1.2 M_{\odot}$.
For fast rotators, the joint effects of radiative levitation and rotational mixing were calibrated by 2D ESTER stellar structure models \citep{Rieutord2013,EspinosaLara2013,Rieutord2016}. This implies that small levels of mixing in the core boundary layers combined with large levels of envelope mixing 
led to better results in forward modelling of $\gamma$ Doradus stars \citep{Mombarg2022}. Just as for internal rotation, $\gamma$ Doradus stars can again be used to downtrack the achieved $D_{\rm mix}(r,t)$ towards stellar models with lower mass for the stars in the P1/P2 sample having a convective core. 

The current sample of asteroseismically modelled $\gamma\,$Doradus stars is insufficient to deduce $D_{\rm mix}(r,t)$ as a good representation of microscopic and macroscopic mixing covering 
($M$, $Z$, $X_C$) in the P1/P2 sample. In particular, almost all of the currently available stars with an estimate of $D_{\rm mix}(r,t)$ are of solar-type metallicity and (too) fast rotation.
Proper calibration of internal mixing due to atomic diffusion, radiative levitation, and rotation altogether requires a set of calibration stars covering a broad range in metallicities and evolutionary stages. For the core program, it is thus required to observe a set of calibrators with $M \geq 1.2 M_{\odot}$ covering $Z \in [0.005,0.04]$, per mass and rotation bin.

The needs for the science calibration of stellar evolution models outlined above give rise to a set of requirements for the sample of $\gamma$ Doradus-type stars: stars of spectral types A9 to F4 and with $P$ magnitudes smaller than 15 will be used to downtrack the rotation and mixing properties in mass to the core science targets. PLATO observations should be conducted through the entire LOPS duration with a cadence of 600 seconds aiming for a precision of 10 to 100 ppm per hour. On-board light curves will provide the necessary quality to achieve these goals.

\subsubsection{{\tt scv5} sample selection}
We use the catalogue by \cite{Hey2024} as our primary source to select candidate $\gamma$ Doradus-type pulsators for observations with the PLATO mission. In their study, the authors revisited a Gaia DR3-based catalog of main-sequence OBAF-type candidate pulsating stars \citep[][]{Gaia2023} to re-classify Gaia light curves with the first two years of TESS photometry. \cite{Hey2024} employed a set of light curves from the TESS Gaia Light Curve (TGLC) catalogue, produced from full-frame images by \cite{Han2023}. All light curves were subjected to frequency analysis using a modification of the pre-whitening algorithm described in \cite{Hey2021}, and a subsequent automated classification according to the type of variability using a feature-based Random Forest classifier.

High-level products of the classification by \cite{Hey2024} are the predicted variability class and the probability of belonging to this class for each star in the sample of $\sim$60,000 considered main-sequence OBAF-type variables. We start our selection process by running the entire sample of $\sim$60,000 classified stars through the PlatoSim tool \citep{Jannsen2024} to select objects that are located within
LOPS2.
In the next step, we extracted all stars that have been classified by \cite{Hey2024} as GDOR/SPB or HYBRID pulsators. The former class represents a group of pure gravity-mode pulsators, where cooler $\gamma$ Doradus-type stars are barely distinguishable from hotter Slowly Pulsating B (SPB)stars from white-light photometry alone. The HYBRID class includes stars in which gravity- and pressure-modes co-exist. Furthermore, we restrict the effective temperature range of the selected gravity-mode and hybrid pulsators to $T_{\rm eff} \in [6000, 10000]$ K to exclude higher mass SPB-type pulsators from the sample. Only stars with a high asteroseismic potential are included in our final selection by rejecting classification outcomes with class probability below 0.5 \citep[see][for a detailed definition of the probabilities]{Hey2024}. This selection turned out to be effective in extracting genuine gravity-mode pulsators. For a subset of those, the internal near-core rotation frequency was estimated by \citet{Aertsetal2025} while their angular momentum was computed and compared with {\it Kepler\/} pulsators by \citet{Aerts2025}. 
Ultimately, to ensure a sufficiently high signal-to-noise ratio is achieved in the PLATO light curves, we select only the pulsators that are brighter than $P_{\rm mag}$ =15. This way, we obtain a sample of 1615 high-priority candidate $\gamma$ Doradus pulsators, among which 225 have an estimate of their near-core rotation frequency. The values of their rotation frequency range from 0.33 per day (corresponding with 3.8$\mu$Hz) until 2.23 per day (25.8$\mu$Hz), ensuring a proper range within the entire population of $\gamma\,$Doradus pulsators \citep[cf.\,Fig.\,1 in][]{Aerts2025}. These pulsators are well characterised in terms of general properties \citep{Mombarg2024a}.

To better meet the requirement for the total number of targets, the sample size has to be increased by almost a factor of three by considering candidate pulsators that are less well characterised asteroseismically so far. We achieve this by further performing an automated classification of all AF-stars found in the LOPS2 field and observed by TESS. Selection of AF-stars is done from a magnitude-limited (down to $G_{\rm mag} = 17$) Gaia stellar catalog and using 2MASS colour-colour cuts as detailed in \cite{IJspeert2024}. From the resulting sample of some 460,000 stars, we select objects with TESS magnitudes ($T_{\rm mag}$) below 13.5. This selection step is dictated by the use in our classification scheme of light curves from the MIT’s Quick-Look Pipeline \citep[QLP,][]{Huang2020}. QLP delivers extracted light curves for objects brighter than $T_{\rm mag}$ = 13.5, and this selection step leaves us with some 160,000 unique objects whose photometric time-series we ultimately run through an automated classification scheme. This scheme is a fully-connected Neural Network (NN) trained on QLP-TESS data of targets that formed the basis of a training set in \cite{Audenaert2021}. We enriched this training set 
with a sample of genuine $\gamma$ Doradus- and $\gamma$ Doradus/$\delta$ Scuti-type hybrid pulsators from \cite{Skarka2022}. We employ wavelet transforms as a set of classification features and achieve ~90\% accuracy on the test set. As in \cite{Hey2024}, we do not stitch individual TESS sectors together into a longer time-series for a given target, instead we classify all sectors individually. This way, we ensure that (in)consistency of the sector-by-sector classification propagates into the final class probability value.

From the results of our automated classification, we select all targets whose TESS sector-based light curves have probability {\tt sector\_prob} $\geq$ 0.9 of belonging to the class of $\gamma$ Doradus-type stars, as well as having a total probability {\tt total\_prob} $\geq$ 0.8. The total probability is computed as a product of the individual, TESS sector-based probabilities and is indicative of the consistency of sector-to-sector classification results. The above selection provides us with a sample of 2700 targets that we assign the same high priority as to the targets selected from the sample of \cite{Hey2024}. In the next step, we lower the probability thresholds to {\tt sector\_prob} $\geq$ 0.85 and 0.7 $\leq$ {\tt total\_prob} $<$ 0.8 to further select 697 second priority candidate $\gamma$ Doradus-type variables.

\begin{figure}[h]
\centering
\includegraphics[width=0.95\textwidth]{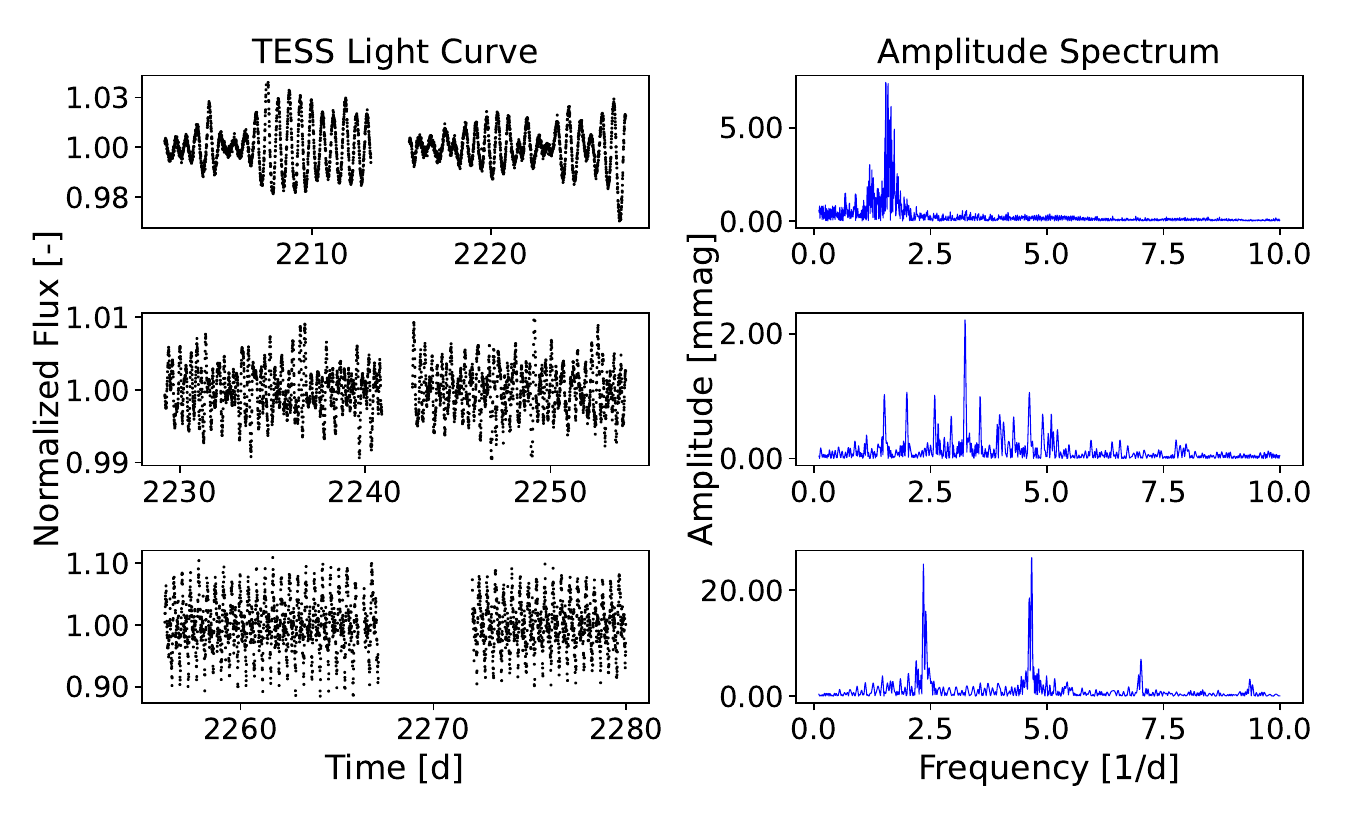}
\caption{Three example TESS light curves (left panels) and amplitude spectra (right panels) of $\gamma$ Doradus pulsators in {\tt scv5}. From top to bottom: TIC 78441682, TIC 320217979, TIC 238664678. }
\label{fig:scv5-examples}
\end{figure}

Altogether, from \cite{Hey2024} and our own additional automated classification, we obtain a selection of 5012 targets, of which 138 are identified as duplicates. Removing the latter, leaves us with the final selection of 4854 priority 1 \& 2 candidate $\gamma$ Doradus-type pulsators. Three examples are shown in Figure \ref{fig:scv5-examples}.

\subsection{Known transiting brown dwarfs ({\tt scv6})}

The intention of {\tt scv6} is to select targets for which precise stellar limb darkening (LD) measurements may be obtained. LD is a critical parameter in the modelling of transiting planetary systems, both for the determination of system parameters and for atmospheric characterisation by transmission spectroscopy \citep[e.g.][]{Me08mn,EspinozaJordan16mn,Coulombe+24aj}. It is also an important phenomenon in the modelling of eclipses in binary star systems and the interferometric determination of stellar angular diameters.

LD is typically implemented using one of several parameterised `laws' of varying utility \citep{Claret00aa,2019A&A...623A.137H}. The LD of stars depends on their atmospheric properties, most strongly effective temperature, and the coefficients of the LD laws can be predicted using model atmospheres. 

Because LD affects the shape of eclipses in binary star and exoplanetary systems, it is possible to measure the LD coefficients by fitting a model to the eclipses in which the coefficients are permitted to vary. For high-quality light curves the LD coefficients can be determined to high precision, although it remains unclear whether this also corresponds to high \emph{accuracy}. These measured LD coefficients can then be compared to theoretical predictions to assess the reliability of the theoretical coefficients. \citet{Me23obs2} showed that all five two-parameter LD laws provide fits to light curves in excellent agreement with each other, suggesting that the choice of LD parameterisation is not critical; however \citet{ClaretSouthworth22aa} found that, of these laws, the power-2 law gives the best match to theoretical LD profiles.

The best objects for measuring LD are eclipsing binaries, preferably with total eclipses, and transiting planets. In both cases the optimal systems are those where the secondary component is much smaller than the primary star, as it blocks only a small fraction of the primary star disc and thus allows a higher spatial resolution of the LD characteristics of the primary star. Bright systems are also favoured due to the high quality of the light curves that can be obtained, and those with longer orbital periods are also better because the components are closer to a spherical shape so proximity effects can be safely neglected.

\subsubsection{{\tt scv6} sample selection}
\label{sec:scv6}

The selection of suitable eclipsing binaries for measuring LD results in a very similar target list as for {\tt scv1a}. So no further eclipsing binaries were added specifically for {\tt scv6}. Similarly, the known transiting planetary systems are already included in the PLATO core science target list \citep{Nascimbeni2025}; so there was no need to include them in {\tt scv6}.
The final category to address is the transiting brown dwarf systems. These are among the most suitable targets due to the small sizes of brown dwarfs compared to their host stars, and are not included in the PLATO core science target list \citep{Nascimbeni2025} because they are too massive to be planets \citep[e.g.][]{Spiegel++11apj,LecavelierLissauer22newar}.

Brown dwarfs are intrinsically rare objects \citep{GretherLineweaver06apj}, but transiting brown dwarfs have been detected as byproducts of photometric searches for transiting hot Jupiters. We queried the TEPCat\footnote{\url{http://www.astro.keele.ac.uk/jkt/tepcat/}} catalogue \citep{Me11mn} to compile a list of all such objects in LOPS2, finding five objects (HIP 33609, TOI-569, TOI-811, TOI-1406 and TOI-2490). One extra object was added in advance of publication (Carmichael, priv.\ comm.), giving a total of six objects. Due to the small number of systems, no attempt was made to define or restrict the statistical properties of these objects.

The five published targets have brown-dwarf masses ranging from 13 to 68 M$_{\rm Jup}$, and orbital periods from 1.9 to 39 d. The host stars have masses of 1.2--2.4 M$_\odot$ and effective temperatures of 5700--10400 K. They thus show quite diverse properties, and allow the effect of LD to be probed over a wide parameter space.


\section{The Science Calibration and Validation PLATO Input Catalogue ({\tt scvPIC})}

The selected {\tt scv} targets are collected in the science calibration and validation PLATO Input Catalog ({\tt scvPIC}), which is a sub-catalogue that gets merged into the PLATO Input Catalogue (PIC). Table \ref{table:LOPS2scvPICtarget2.2.0.2} shows the number of targets in each {\tt scv} sample of the {\tt scvPIC} catalog version LOPS2scvPICtarget2.2.0.2. The total count of targets in the {\tt scvPIC} is 38,585, with some targets attributed to multiple {\tt scv} samples.

Some targets in the {\tt scvPIC} are known binaries from Gaia DR3 and are identified as non-single star systems via the NSSflag.
Some of these known binaries (e.g., eclipsing binaries) appear as a single target in the {\tt scvPIC} while others of these known binaries (e.g., wide binaries) appear as multiple targets in the {\tt scvPIC}. 
If targets were not labelled as binaries in Gaia DR3 (e.g., because recent observations revealed their binarity), they are not identified as such in the {\tt scvPIC}. 

The generation of the {\tt scvPIC} starts with the input list that is compiled by the team of {\tt scvPIC} scientists under coordination of the PLATO Science Management (PSM) as described in Section \ref{sec:scvtargets}. This {\tt scvPIC} input list contains four columns:

\begin{itemize}
  \item target ID: star name provided by the {\tt scvPIC} scientists, preferably Gaia DR3 ID
  \item {\tt scvPICsourceFlag}: identification of {\tt scv} sample
  \item $G$: Gaia $G$-band magnitude from Gaia DR3
  \item priority metric:
  \vspace{-2mm}
  \begin{itemize}
  \setlength
    \item[-] priority 1: target crucial for the {\tt scv} sample to achieve the corresponding goal
    \item[-] priority 2: target extends the parameter space for calibration purposes
    \item[-] priority 3: any other target that fits the {\tt scv} sample definition
  \end{itemize}
\end{itemize}

\begin{figure}[h]
\centering
\includegraphics[width=0.8\textwidth]{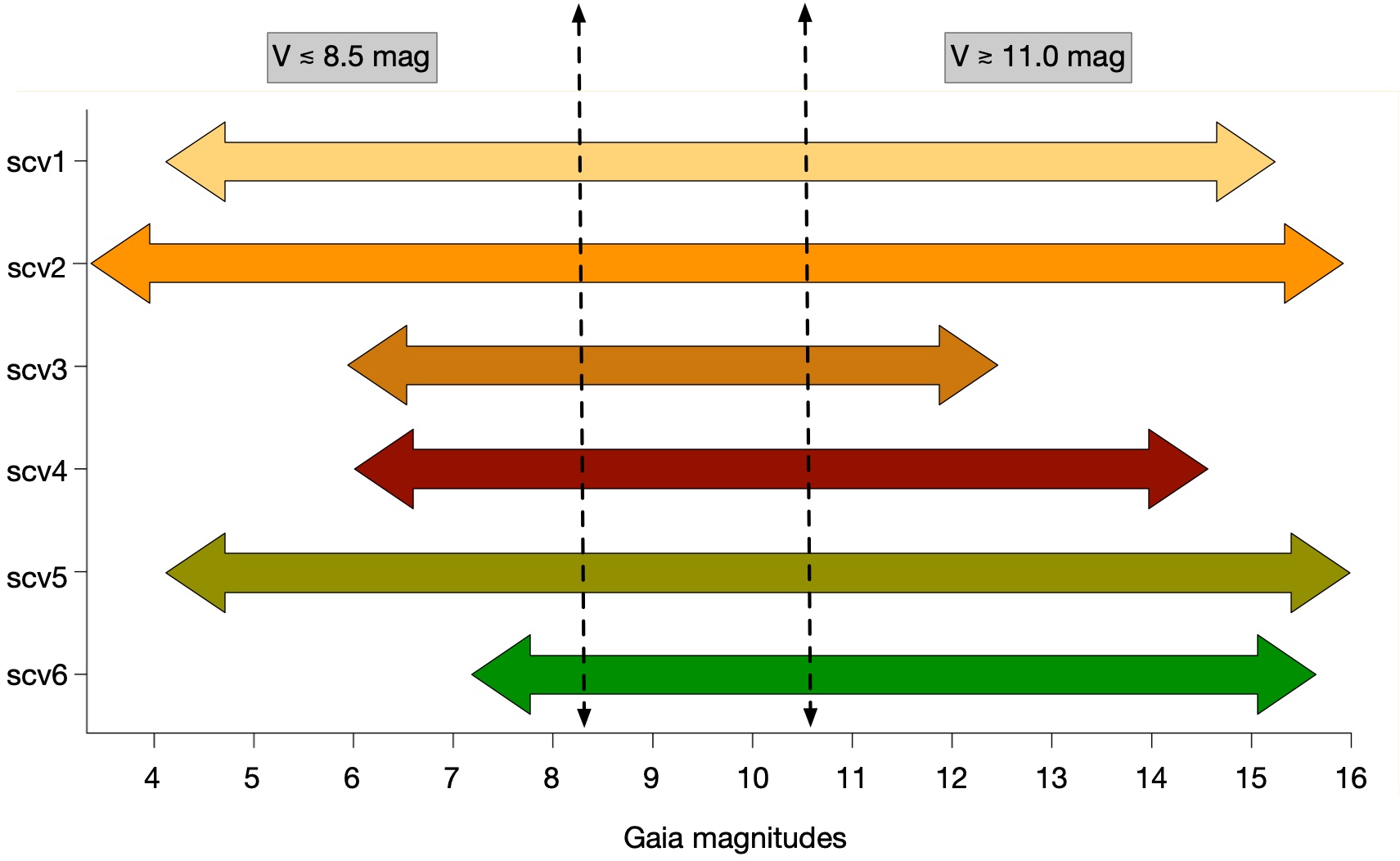}
\caption{Gaia $G$ band magnitude ranges for each of the six {\tt scv} samples. The vertical dashed lines mark the positions of $V$ magnitudes 8.5 and 11 mag that approximately correspond to average $G$ magnitudes in the {\tt scvPIC} sample of 8.3 and 10.6, respectively.  }
\label{fig:scv-magnitudes}
\end{figure}

The Gaia $G$-band magnitudes are in the range from 4.1 to 15.2 mag for the {\tt scv1} targets, from 3.4 to 15.9 mag for the {\tt scv2} targets, from 6.0 to 12.4 mag for the {\tt scv3} targets, from 6.1 to 14.5 mag for the {\tt scv4} targets, from 4.2 to 15.1 mag for the {\tt scv5} targets, and from 7.2 to 15.6 mag for the {\tt scv6} targets. Figure \ref{fig:scv-magnitudes} illustrates the magnitude ranges for each {\tt scvPIC} sample. The positions of the $V$ magnitude values 8.5 and 11 mag marked in Figure \ref{fig:scv-magnitudes} as vertical dashed lines were derived by a statistical comparison of $V$ and $G$ magnitudes in the {\tt scvPIC} sample and were found to correspond on average to approximately $G = 8.3$ and 10.6 mag. 

The resulting input list (LOPS2scvPICtargetNSRinput2.2.0.1.csv) was delivered\footnote{on 16 June 2025} to the PLATO Data Center Data Base (PDC-DB). The PDC then checks the list for targets that appear in multiple {\tt scv} samples. Any multiple entries are merged into the {\tt scvPICsourceFlag}, which is a bitmask indicating which {\tt scvPIC} sample an object belongs to. This way, potential multiple {\tt scv} sample memberships for a target are preserved and the final {\tt scvPIC} contains only a single entry for each target.

PDC then generates the {\tt scvPIC}, which follows a pre-defined data model with 71 columns for each target. These columns contain, for example, the right ascension and declination of the target in the International Celestial Reference System (ICRS)\footnote{If the proper motion is available from Gaia DR3, then the coordinates are predicted for the expected middle-epoch of the first LOP field observation (2028.3). If the proper motion is unavailable, then the epoch of the coordinates is set to the reference epoch of the right ascension and declination in the corresponding source catalog, for example about 1991 for Tycho (details depending on the specific target) and 2016.0 for Gaia DR3.}, the proper motion at the reference epoch of the source catalog\footnote{The reference epoch is 2016.0 if the proper motion is from Gaia DR3, but the reference epoch may differ for other source catalogs.}, the estimated PLATO magnitudes in the N-CAM and in the F-CAM, stellar effective temperature etc. Most of this information, including the `star name' identifier, is taken from a common list of contaminants of the PIC shared by the PDC \citep{Marrese2026}, which contains about 100,000,000 Gaia DR3 targets in LOPS2 plus a few bright targets in LOPS2 that are not in Gaia DR3. All {\tt scvPIC} targets can be found in this list of contaminants and their data from this list is used in the {\tt scvPIC}\footnote{There are nine targets from Hipparcos and six from Tycho in the common list of contaminants shared by the PDC (LOPS2PICcontaminantMPSSR2.2.0.1). As a result, not all the info of {\tt scvPIC} are from Gaia.}. More generally, the list of PIC contaminants includes all the targets of the four PIC sub-catalogues (tPIC, scvPIC, cPIC, and fgPIC) by construction, because these subPICs are generated based on the list of contaminants \citep{Heller2026,Marrese2026}.

The signal and noise budget for the {\tt scvPIC} targets is computed using the approach described in \cite{Boerner2024}. The instrument model describes the signal flow from the incoming stellar flux star to a digital output considering the main optical, mechanical, thermal and electrical effects, including all known noise sources. The computation of the signal and noise budget for {\tt scvPIC} targets is analogue to that of {\tt tPIC} targets.

PDC then performs several checks as part of the technical verification of the data product, such as the above-mentioned merge of multiple entries, attribution of unique {\tt scvPIC} IDs for each {\tt scvPIC} target, a comparison of the submitted target IDs from the {\tt scvPIC} scientists and the selected star name identifier, and visualizations of all {\tt scv} samples in the LOPS2 field to aid debugging and to identify any outliers.





\begingroup

\setlength{\tabcolsep}{2.5pt} 
\renewcommand{\arraystretch}{1.5} 

\begin{table}[h]
\caption{Number count of the {\tt scvPIC} samples in the {\tt scvPIC} catalog version LOPS2scvPICtarget2.2.0.2.
}
\label{table:LOPS2scvPICtarget2.2.0.2}
\centering
\begin{tabular}{c|c|c|c|c|c|c|c|c|c|c|c|c|c|c}
\toprule
 & {\tt 1a} & {\tt 1b} & {\tt 1c} & {\tt 1d} & {\tt 1e} & {\tt 2a} & {\tt 2b} & {\tt 3a} & {\tt 3b} & {\tt 4a} & {\tt 4b} & {\tt 4c} & {\tt 5} & {\tt 6}  \\ 
\midrule
 all & 849 & 1539  &  806  &   1   &  53   &   50  &  306  &   54  &  173  &   233   &   1592    & 28175 & 4854 & 6  \\
 \midrule
$V < 8.5$ & 18 & 70  &  59  &   0   &  9   &   45  &  64  &   6  &  3  &   2   &   5    & 173 & 81 & 0 \\
$8.5 < V < 11$ & 205 & 828  &  185  &   0   &  44   &   3  &  179  &   19  &  19  &   13   &   39    & 2879 & 902 & 1 \\
$V > 11$ & 477 & 641  &  562  &   0   &  0   &   2  &  63  &   2  &  0  &   218   &   1543    & 25123 & 3293 & 4 \\
no $V$ & 149 & 0  &  0  &   1   &  0   &   0  &  0  &   27  &  151  &   0   &   5    & 0 & 578 & 1 \\
\botrule
\end{tabular}
\footnotetext{The top line (all) gives full target numbers per {\tt scvPIC} subsample. The subsequent three lines list the numbers of {\tt scvPIC} targets in the magnitude ranges brighter than $V = 8.5$ mag ($V<8.5$), between 8.5 and 11 mag in $V$ ($8.5 < V < 11$), and fainter than $V=11$ mag ($V>11$). The last line (no $V$) lists the number of targets for which no $V$ magnitudes are available.}
\end{table}

\endgroup

\begin{figure}[h]
\centering
\includegraphics[width=0.9\textwidth]{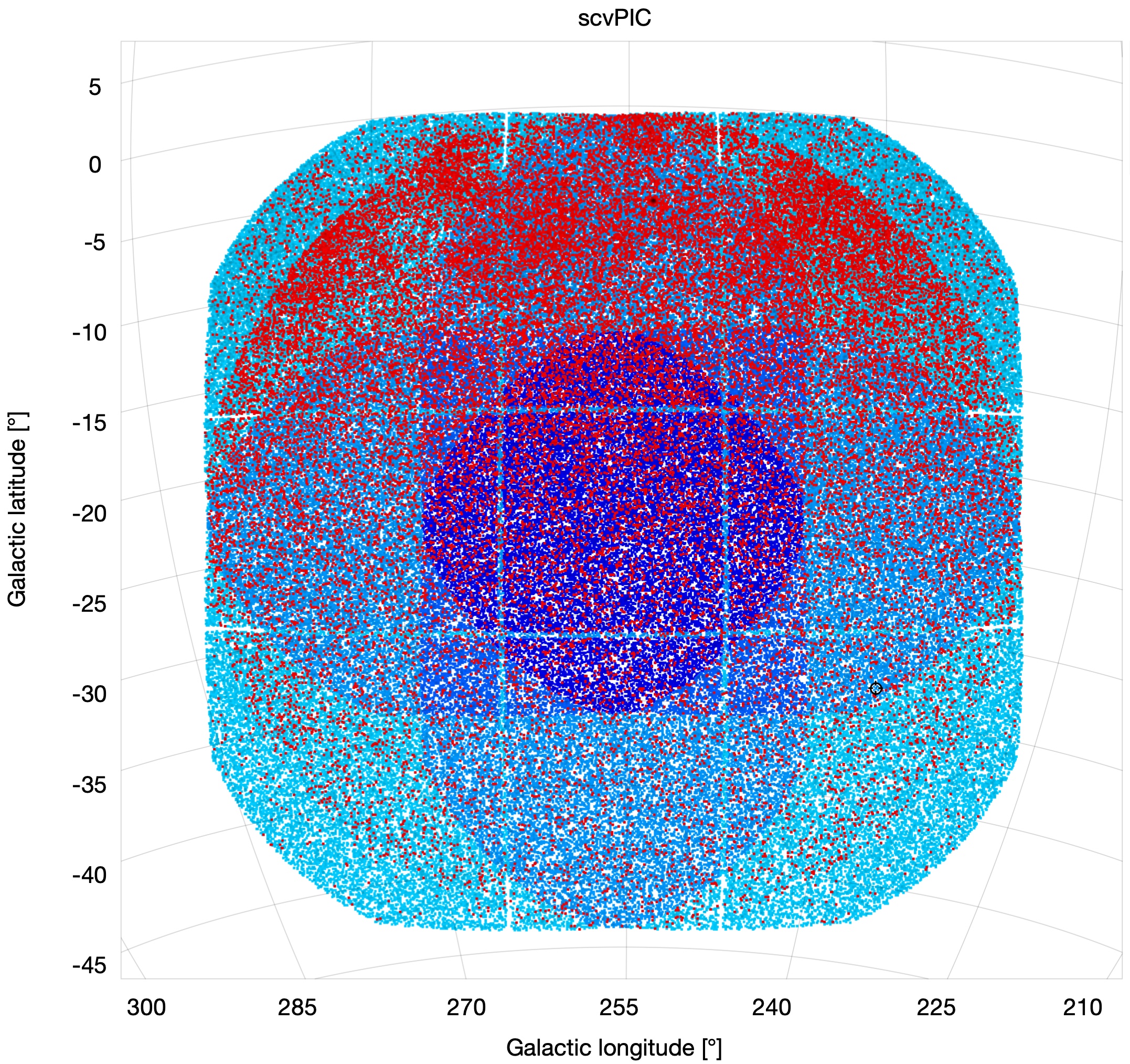}
\caption{The location of the {\tt scvPIC} targets in LOPS2 (red points; based on PIC v2.2.0.1). We only show those 38,007 {\tt scvPIC} targets that are predicted to be observed by any of the PLATO N-CAMs. Light, medium, darker, and darkest blue colours show the areas monitored by 6, 12, 18 and 24 normal cameras (N-CAMs), respectively. The blue points are all targets in the PIC making up the `PLATO footprint'.}
\label{fig:scvpic-sky}
\end{figure}

Figure \ref{fig:scvpic-sky} shows the location of all {\tt scvPIC} targets in LOPS2.

\begin{figure}[h]
\centering
\includegraphics[width=0.9\textwidth]{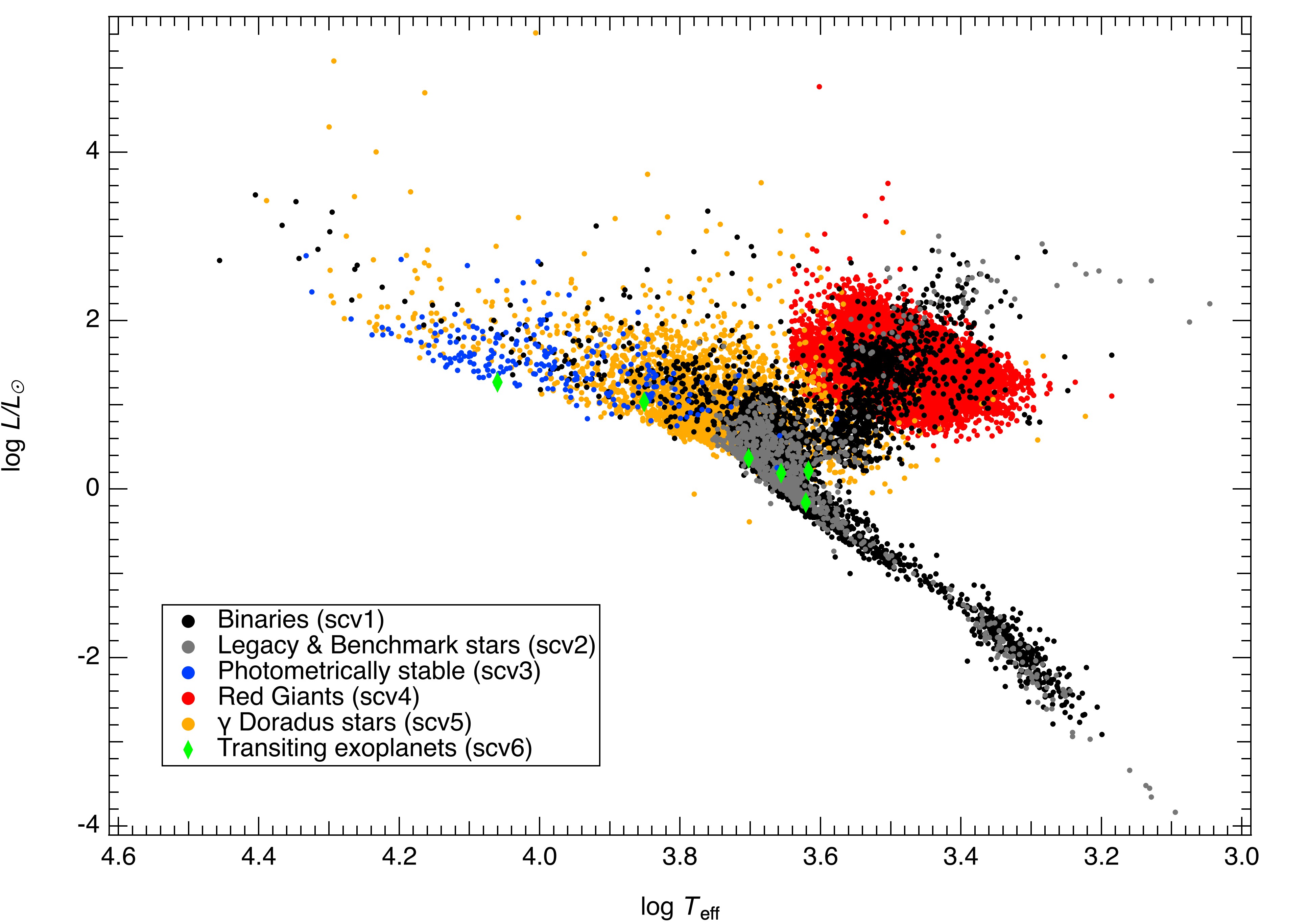}
\caption{The {\tt scvPIC} targets in the HR-diagram. The values for $\log T_{\rm eff}$ and $\log L/L_{\odot}$ were derived from Gaia photometry.}
\label{fig:scvpic-hrd}
\end{figure}

The {\tt scvPIC} targets include the evolutionary stages from the main sequence to the red giant phases as can be seen in the Hertzsprung-Russell diagram in Figure \ref{fig:scvpic-hrd}.

The final data products for the {\tt scvPIC} targets are still under review. By default on-board light curves will be observed for all targets except for those brighter than $V =8.5$ mag to avoid saturation. 
535 of the {\tt scvPIC} targets are brighter than $V =8.5$ mag (or approximately $G = 8.3$ mag, see Figure \ref{fig:scv-magnitudes}) and are therefore proposed to be observed as imagettes. 156 of those are in {\tt scv1}, 109 in {\tt scv2}, 9 in {\tt scv3}, 180 in {\tt scv4}, 81 in {\tt scv5} and none in {\tt scv6} (see Table \ref{table:LOPS2scvPICtarget2.2.0.2}). We also strive to obtain imagettes for the {\tt scvPIC} targets in clusters as cluster members are key calibrators for stellar evolution and are typically located in crowded regions which is why imagettes are required.


\section{Conclusion}

In this paper, we outlined the PLATO science calibration and
validation plan. More than 38,500 targets in the first long-pointing
field of the mission, LOPS2 in the southern hemisphere, were included
in the {\tt scvPIC} to achieve the goals of this plan. These {\tt
scvPIC} targets were added to the overall LOPS2 PIC, which includes
the targets of the core science program of PLATO ({\tt tPIC}), as well
as the fine guidance stars ({\tt fgPIC}) to regulate the attitude
control of the spacecraft and the instrument calibration stars ({\tt cPIC}). 
The various sub-PICs each have their own
specific selection function, as described in other accompanying papers \citep[][Nascimbeni et al. 2026, submitted; Heller et al. 2026, submitted to {\it Experimental Astrononmy}]{Nascimbeni2022,Nascimbeni2025,Prisinzano2026}.

Following the stringent requirement of achieving 10\% age accuracy
from the stellar models to be used in the asteroseismic pipeline of
the mission for the dwarfs and subgiants in PLATO's prime sample
(Nascimbeni et al., submitted), the two largest sub-samples of the
{\tt scvPIC} consist of hydrogen-shell and helium-core burning red
giants and core-hydrogen burning intermediate-mass gravity-mode
pulsators. These tens of thousands of pulsators in the {\tt scvPIC}
were carefully selected by relying on their asteroseismic properties
observed by the TESS mission.

The {\tt scvPIC} sample of stars was composed following these stars'
optimal potential to validate exoplanetary transits and to improve the
theory of stellar interiors, notably in terms of angular momentum
transport and internal mixing. For optimal science calibration of
transport processes inside the stars, the candidate pulsating
{\tt scvPIC} stars were chosen such as to cover both the hydrogen and
helium nuclear burning stages.

Overall, the {\tt scvPIC} is composed of targets covering a
broad range of masses, metallicies, evolutionary stages, and rotation
velocities so as to make it representative of stars in our close
vicinity in the Milky Way.  Moreover, known systems with transiting
exoplanets and particularly well-known benchmark, binary, and
photometrically stable stars were included to validate the analysis
pipelines from in-orbit data for these objects. The selection was done
such as to achieve the overall best performance of the mission,
including optimal age-dating of the stars in the prime sample.

Following general ESA policies, the PLATO in-orbit measurements of all the {\tt scvPIC} targets will be made publicly available, once properly corrected for instrumental effects and carefully validated by the instrument teams.

\bmhead{Acknowledgements}
This publication is dedicated to the late Patrick Gaulme, who contributed actively to the preparation of the {\tt scvPIC}, but passed away at far too young age in July 2025.
This work presents results from the European Space Agency (ESA) space mission
PLATO. The PLATO payload, the PLATO Ground Segment and PLATO data processing
are joint developments of ESA and the PLATO Mission Consortium (PMC). Funding for
the PMC is provided at national levels, in particular by countries participating in the
PLATO Multilateral Agreement (Austria, Belgium, Czech Republic, Denmark, France,
Germany, Italy, Netherlands, Portugal, Spain, Sweden, Switzerland, Norway, and United
Kingdom) and institutions from Brazil. Members of the PLATO Consortium can be found
at \url{https://platomission.com/}. The ESA PLATO mission website is
\url{https://www.cosmos.esa.int/plato}. We thank the teams working for PLATO for all their work.\\
C. Aerts, A.\ Tkachenko, P.\ Huijse, N. Jannsen, and T.\ Morel acknowledge financial support from the Belgian Science Policy Office 
Belspo, PRODEX contracts for PLATO mission development and Gaia data exploitation.
M.\ Kliapets is supported by a Kavli Foundation Scholarship.
A. Lanza and J. Montalban acknowledge Accordo ASI-INAF n. 2022-28-HH.0 fase D progetto PLATO. 
P. Maxted acknowledges support from UK Science and Technology Facilities Council (STFC) research grant numbers
UKRI1193 and ST/Y002563/1 and UK Space Agency (UKSA) grant number UKRI966.
L. Briganti aknowledges financial support from MUR (Ministero dell'Università e della Ricerca) and NextGenerationEU throughout the PNRR ex D.M. 118/2023.
A. Miglio acknowledges support from the ERC  Consolidator Grant funding scheme (project ASTEROCHRONOMETRY, G.A. n. 772293).
S. Sulis acknowledges support from CNES.
R. Heller, C. Jiang, and M. Ammler-von Eiff acknowledge support from the German Aerospace Agency (Deutsches Zentrum f{\"u}r Luft- und Raumfahrt) under PLATO Data Center grants 50OO1501 and 50OP1902.
J. M. Mas-Hesse is funded by Spanish MICIU/AEI/10.13039/501100011033 and ERDF/EU grant PID2023-147338NB-C21.
U. Heiter and N. Miller acknowledge support from the Swedish National Space Agency (SNSA/Rymdstyrelsen).
K. G. Helminiak is supported by the Polish National Science Center through grant no. 2023/49/B/ST9/01671.
O. Kochukhov acknowledges support by the Swedish Research Council (grant agreement 2023-03667) and the Swedish National Space Agency.
M. N. Lund acknowledges support from the ESA PRODEX programme (PEA 4000142995).
T. Merle is granted by the BELSPO Belgian federal research program FED-tWIN under the research profile Prf-2020-033\_BISTRO. 
D.B. Palakkatharappil acknowledges the support from the PLATO Centre National D'{\'E}tudes Spatiales grant at CEA.
J. Southworth acknowledges support from STFC under grant number ST/Y002563/1.
A. Triaud is supported by a grant from the European Research Council (ERC) under the European Union's Horizon 2020 research and innovation programme (grant agreement n$^\circ$ 803193/BEBOP) and by the ERC/UKRI Frontier Research Guarantee programme (EP/Z000327/1/CandY).
M. Vrard acknowledges funding from the European Research Council (ERC) under the European Union’s Horizon 2020 research and innovation programme (G.A. n. 10101965).
P. G. Beck acknowledges support by the Spanish Ministry of Science and Innovation with the \textit{Ram{\'o}n\,y\,Cajal} fellowship number RYC-2021-033137-I, as well as the proyecto plan nacional \textit{PLAtoSOnG} (grant no. PID2023-146453NB-100).

\section*{Declarations}
\begin{itemize}
\item Funding: There is no funding. 
\item Competing interests: The authors declare no competing interests.
\item Data availability: Data presented in this manuscript are included in the catalogue PIC v2.2.0.1 which is publicly available. 
\item Ethics declaration: Not applicable.
\end{itemize}


\begin{appendices}

\section{Location of the {\tt scv} samples within the LOPS2 field}\label{secA1}

The location of the {\tt scvPIC} targets per subsample (based on catalogue version PIC v2.2.0.1) in LOPS2.

\begin{figure}[h]
\centering
\includegraphics[width=0.49\textwidth]{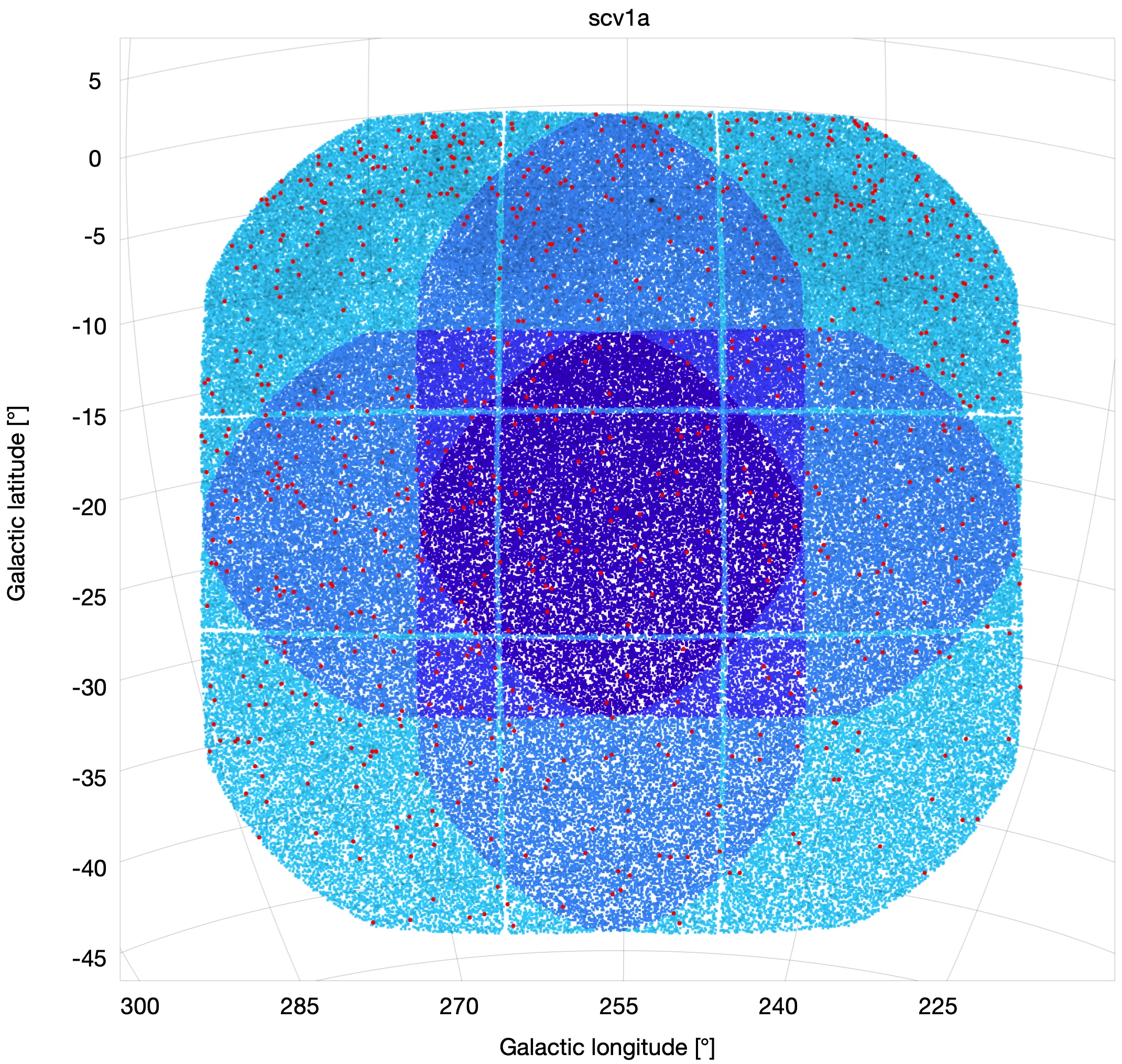}
\includegraphics[width=0.49\textwidth]{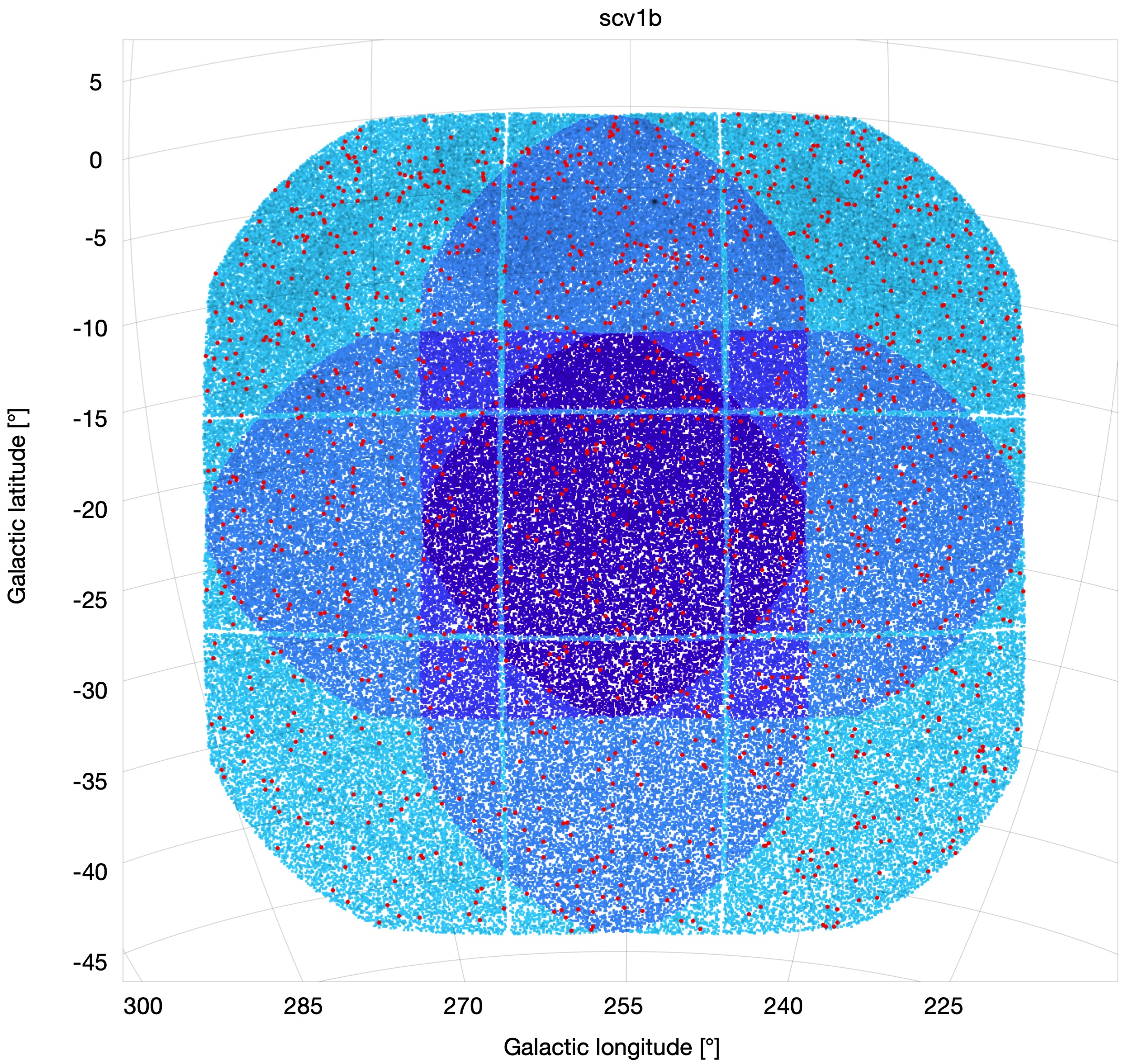}
\caption{Positions of the {\tt scv1a} (left) and {\tt scv1b} (right) targets in LOPS2. Colour-code as in Figure \ref{fig:scvpic-sky}.}
\label{fig:LOPS_scv1ab}
\end{figure}

\begin{figure}[h]
\centering
\includegraphics[width=0.49\textwidth]{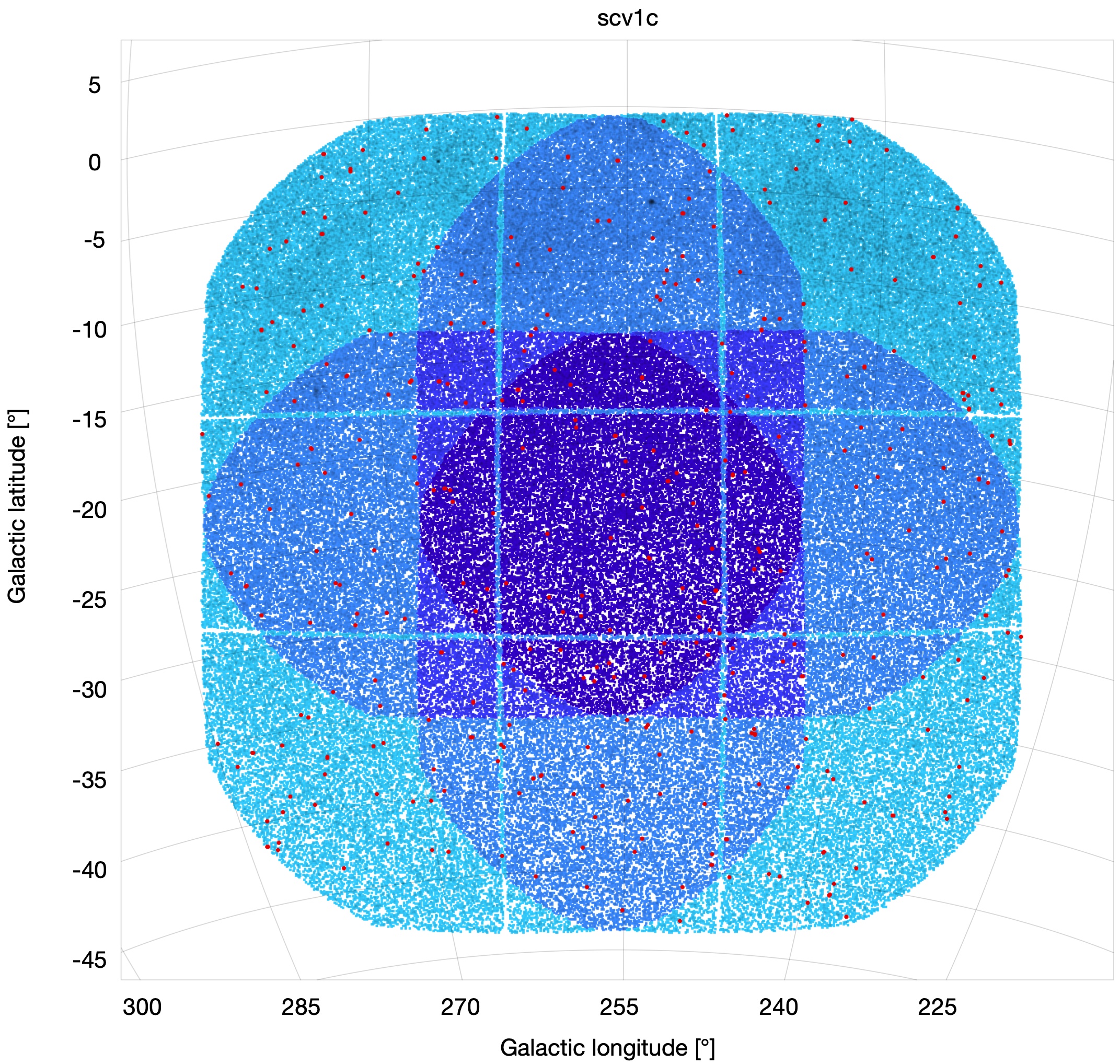}
\includegraphics[width=0.49\textwidth]{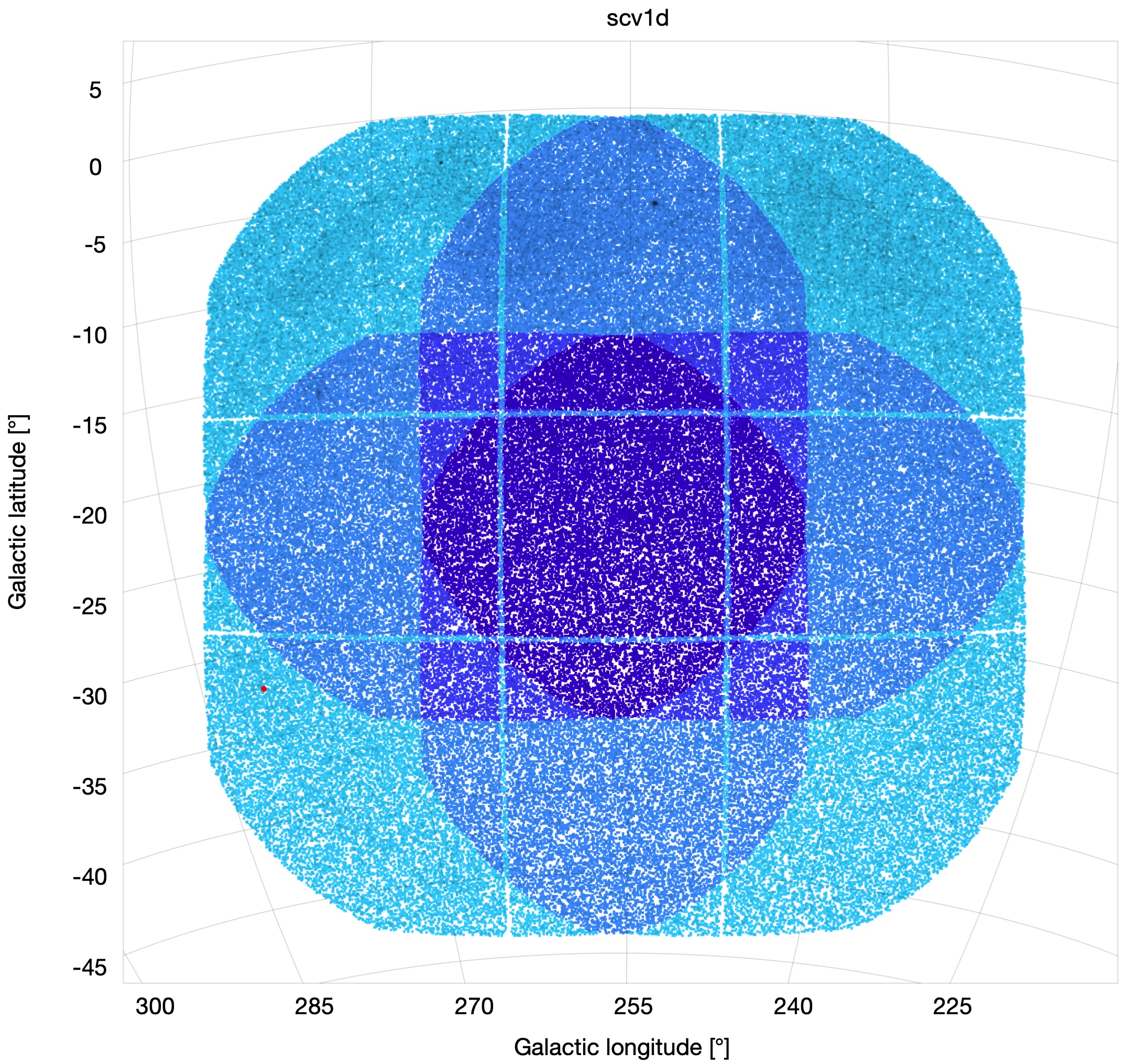}
\caption{Positions of the {\tt scv1c} (left) and {\tt scv1d} (right) targets in LOPS2. Colour-code as in Figure \ref{fig:scvpic-sky}.}
\label{fig:LOPS_scv1cd}
\end{figure}

\begin{figure}[h]
\centering
\includegraphics[width=0.49\textwidth]{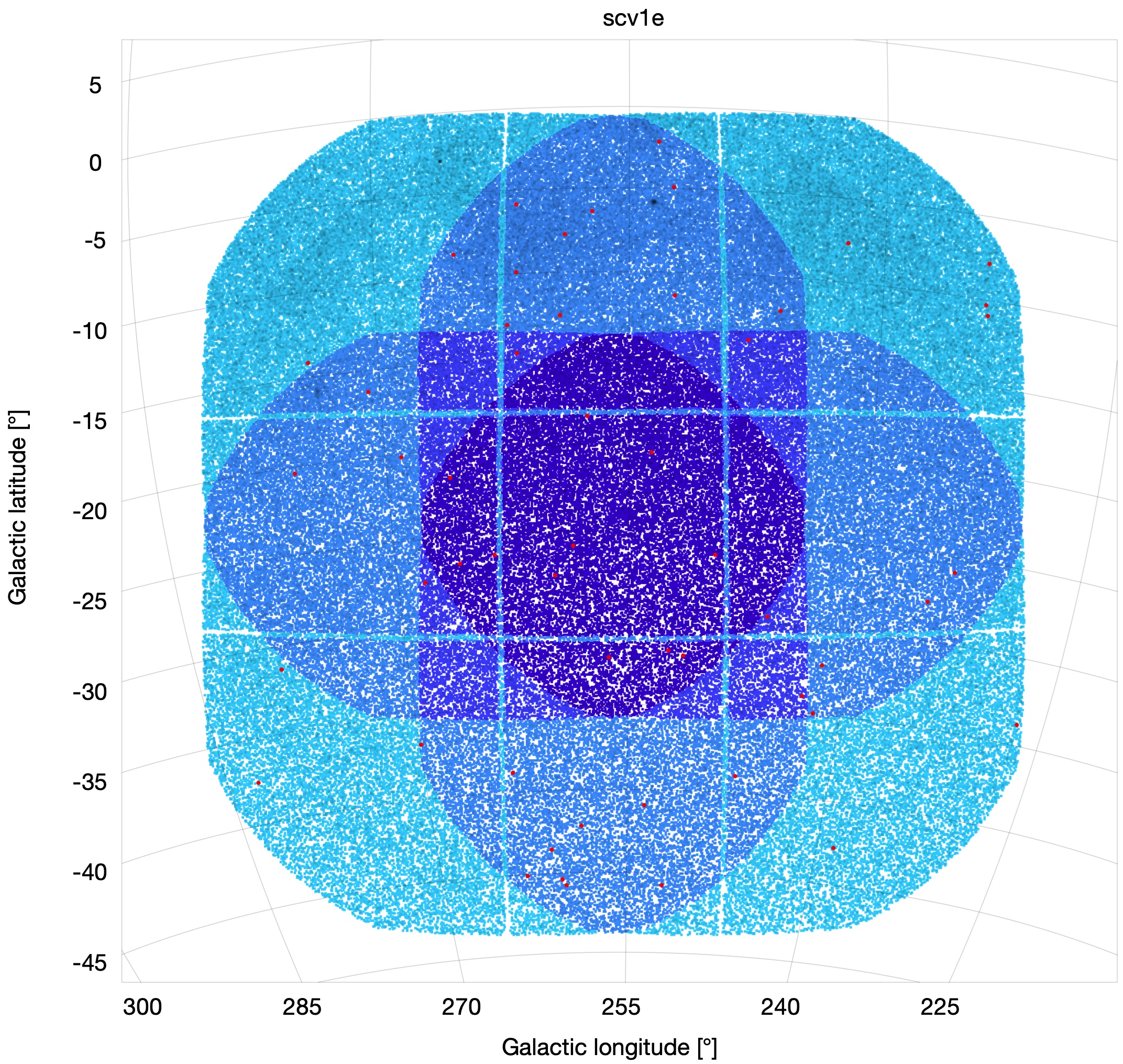}
\caption{Position of the {\tt scv1e} targets in LOPS2. Colour-code as in Figure \ref{fig:scvpic-sky}.}
\label{fig:LOPS_scv1e}
\end{figure}

\begin{figure}[h]
\centering
\includegraphics[width=0.49\textwidth]{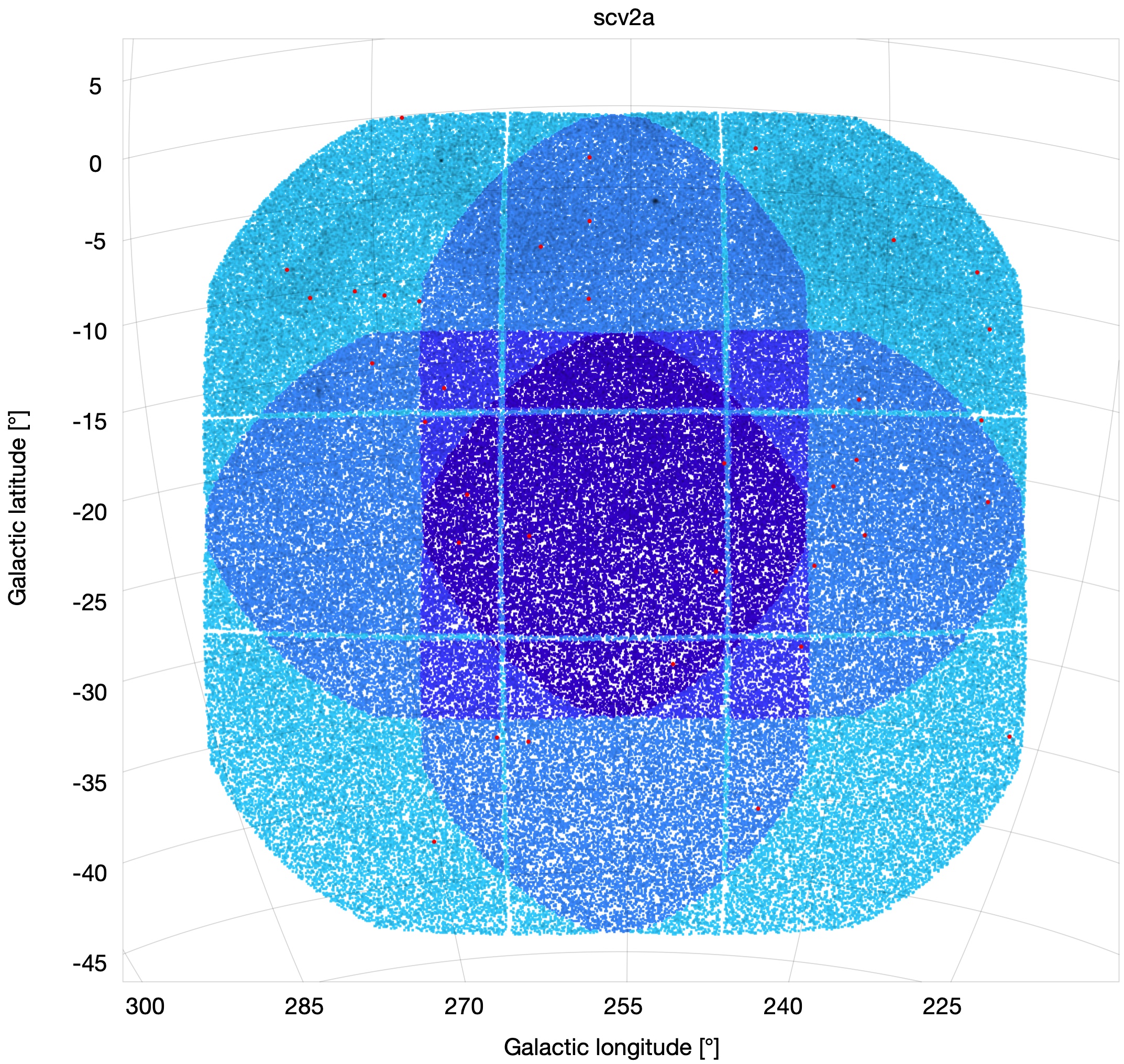}
\includegraphics[width=0.49\textwidth]{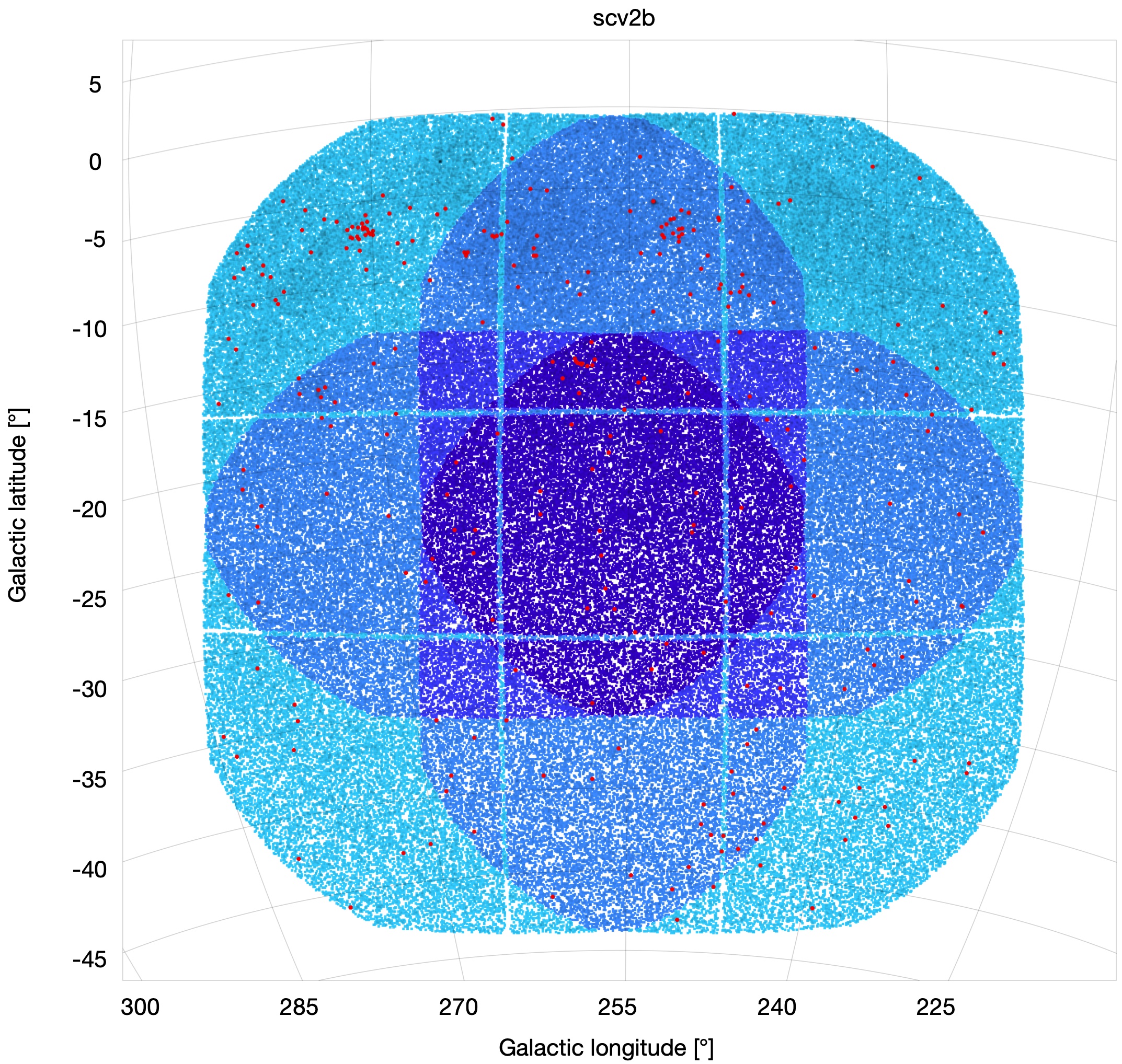}
\caption{Positions of the {\tt scv2a} (left) and {\tt scv2b} (right) targets in LOPS2. Colour-code as in Figure \ref{fig:scvpic-sky}.}
\label{fig:LOPS_scv2ab}
\end{figure}

\begin{figure}[h]
\centering
\includegraphics[width=0.49\textwidth]{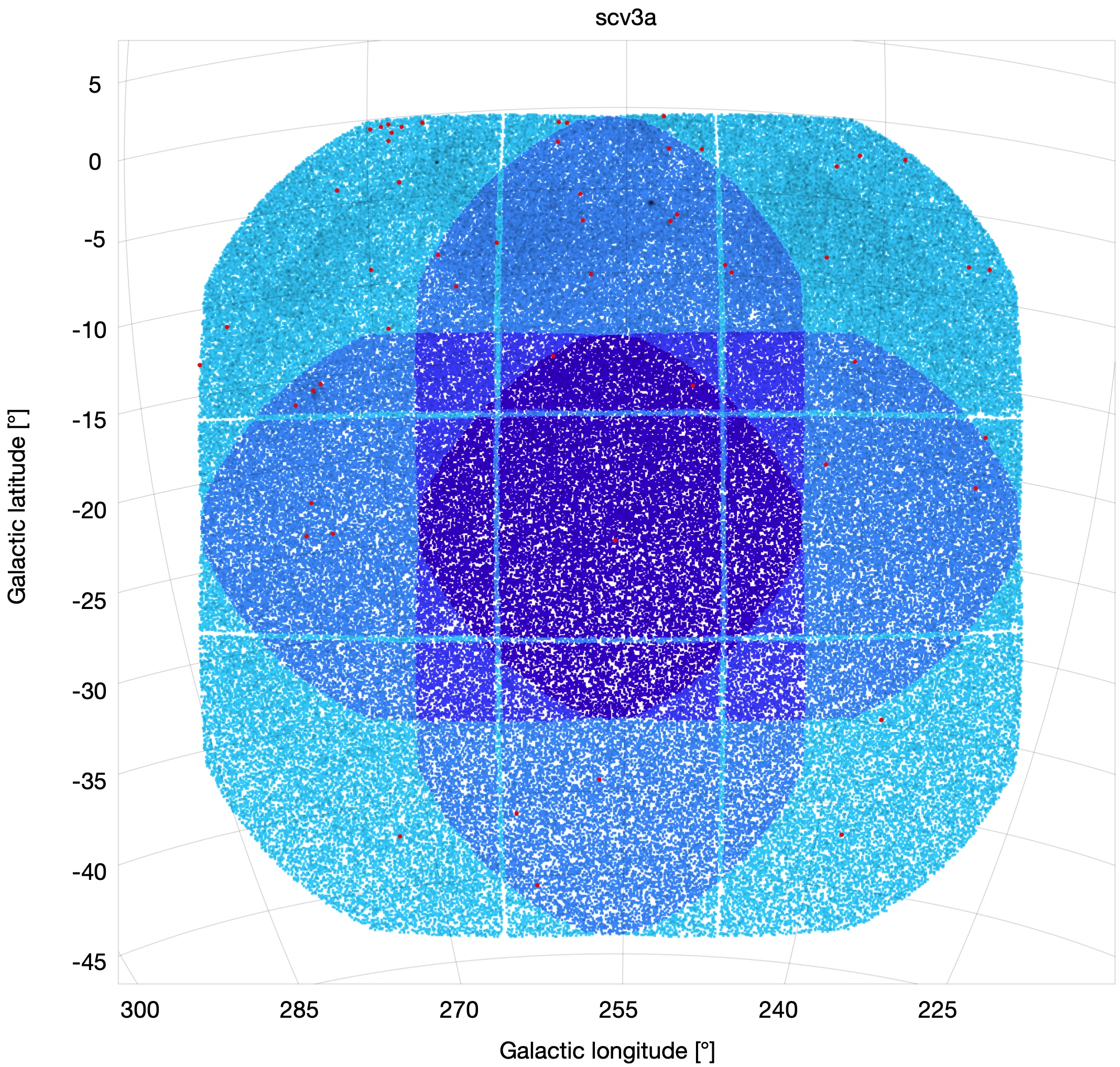}
\includegraphics[width=0.49\textwidth]{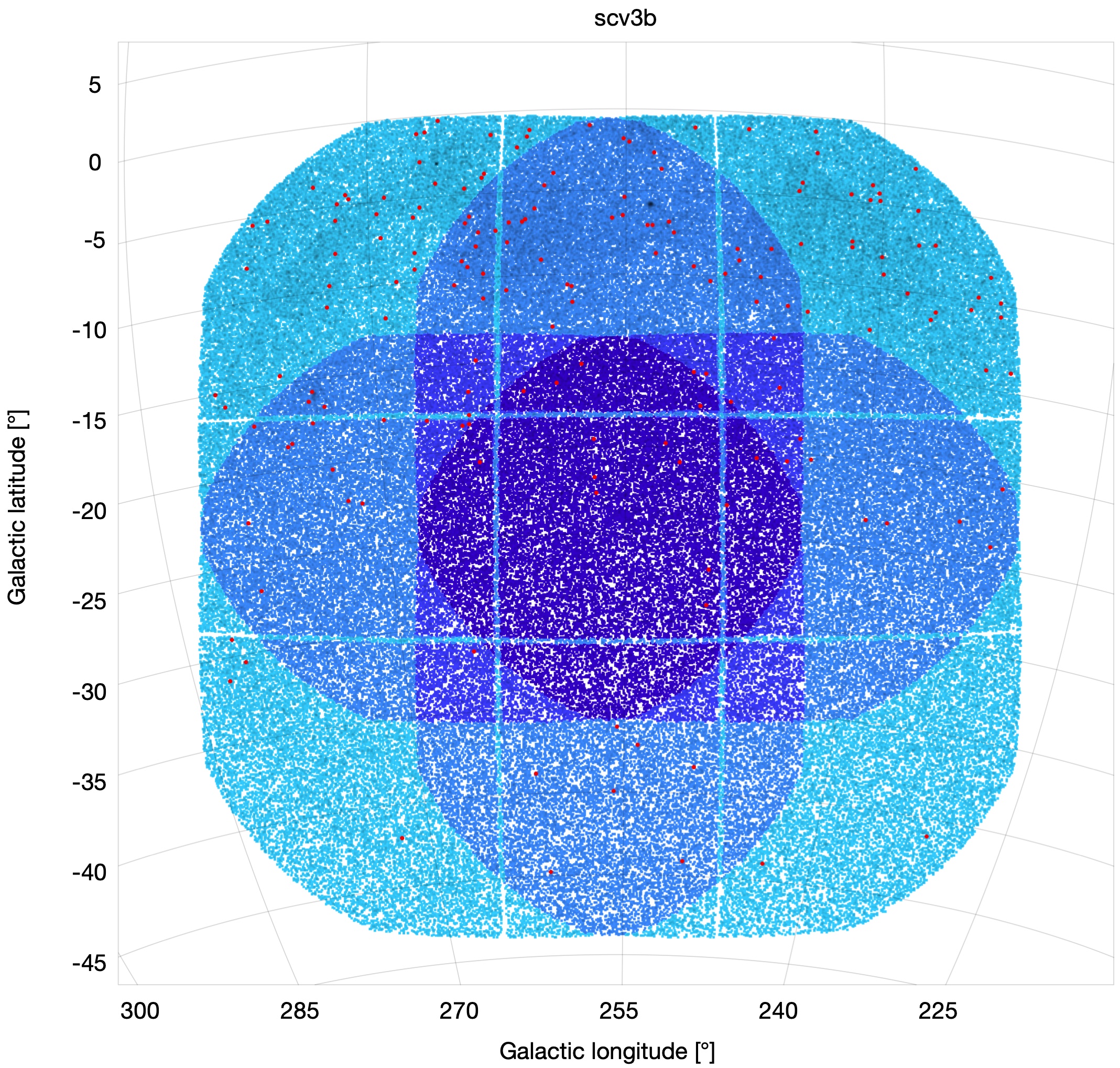}
\caption{Positions of the {\tt scv3a} (left) and {\tt scv3b} (right) targets in LOPS2. Colour-code as in Figure \ref{fig:scvpic-sky}.}
\label{fig:LOPS_scv3ab}
\end{figure}

\begin{figure}[h]
\centering
\includegraphics[width=0.49\textwidth]{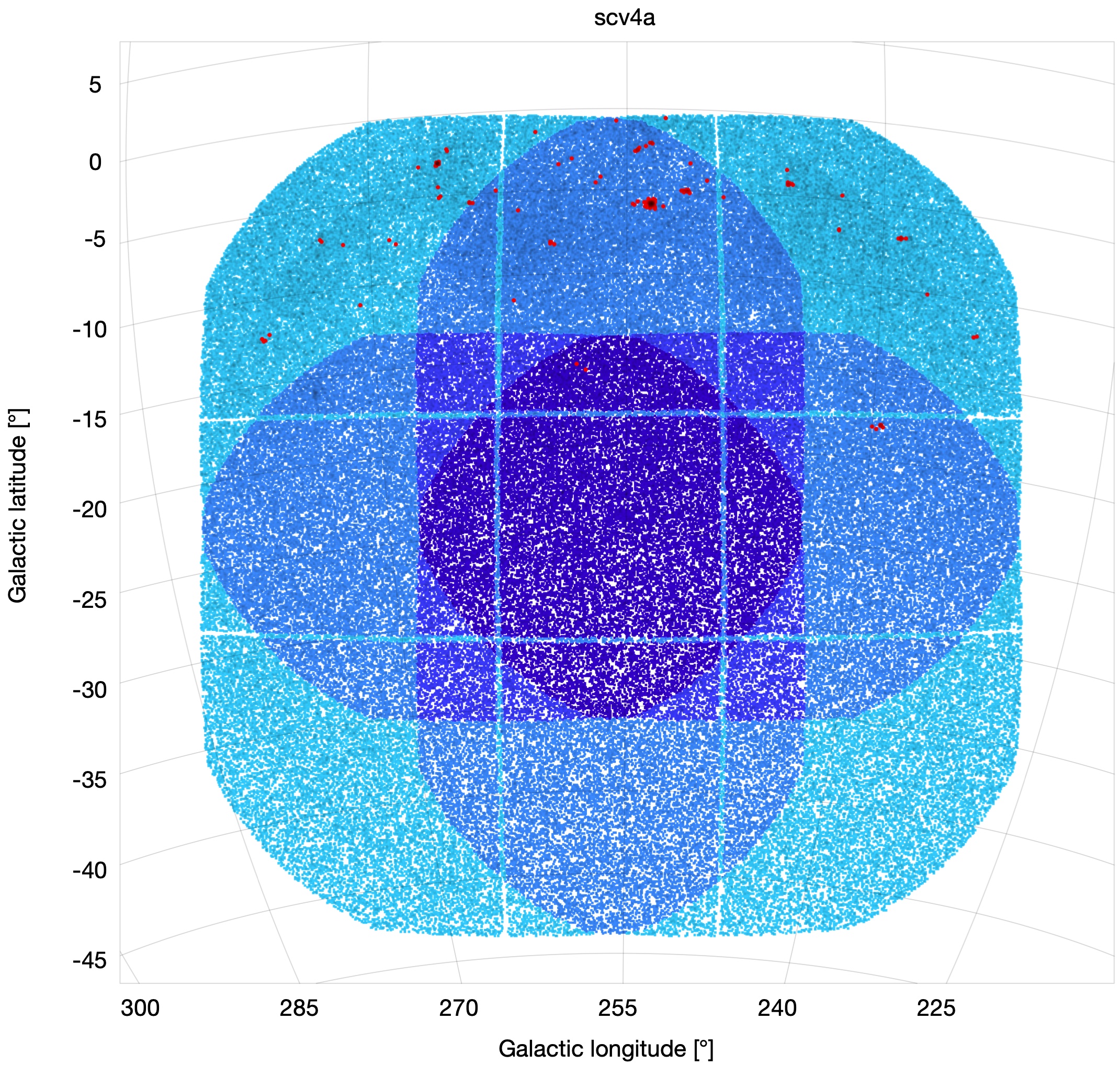}
\includegraphics[width=0.49\textwidth]{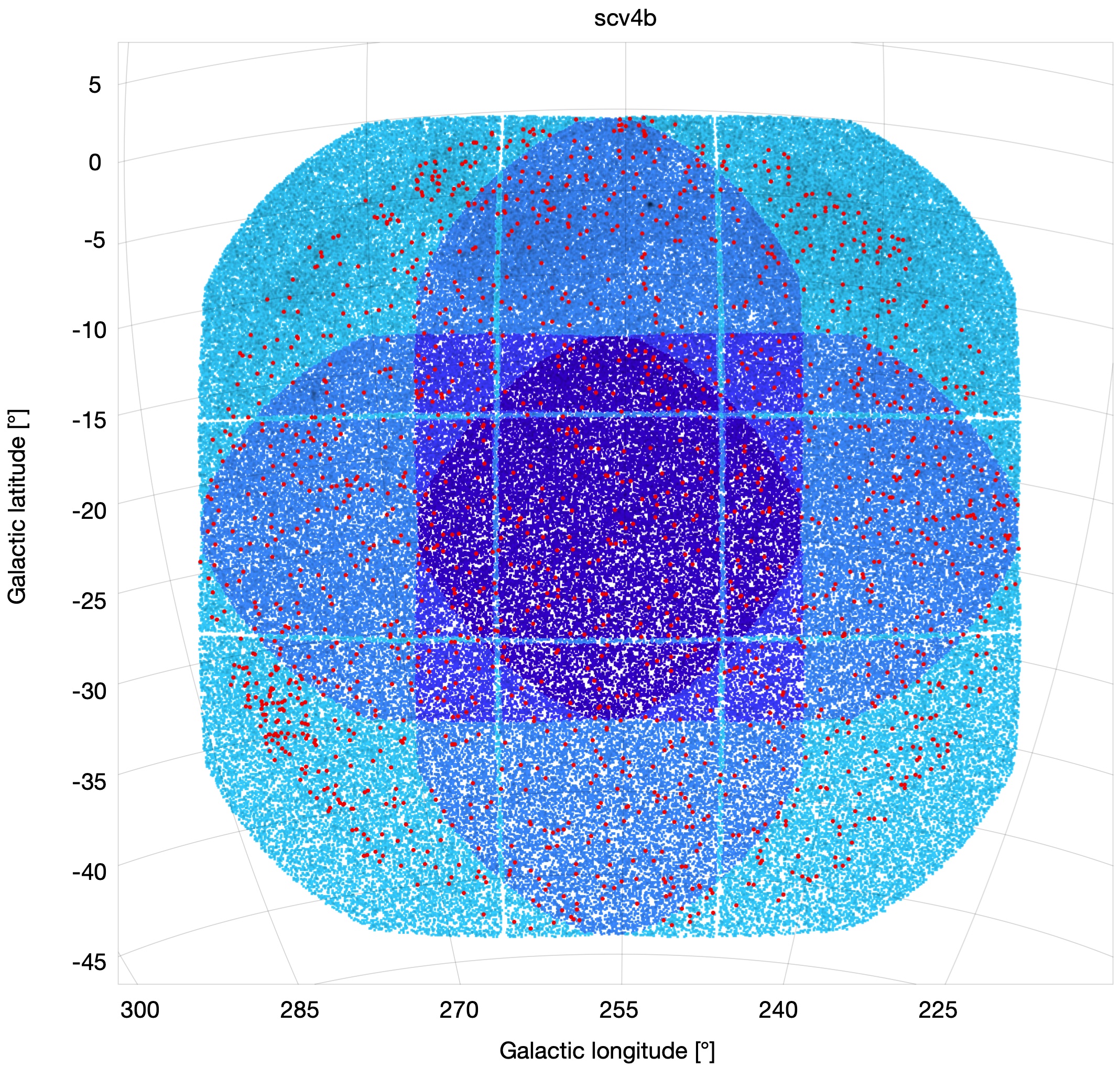}
\caption{Positions of the {\tt scv4a} (left) and {\tt scv4b} (right) targets in LOPS2. Colour-code as in Figure \ref{fig:scvpic-sky}.}
\label{fig:LOPS_scv4ab}
\end{figure}

\begin{figure}[h]
\centering
\includegraphics[width=0.49\textwidth]{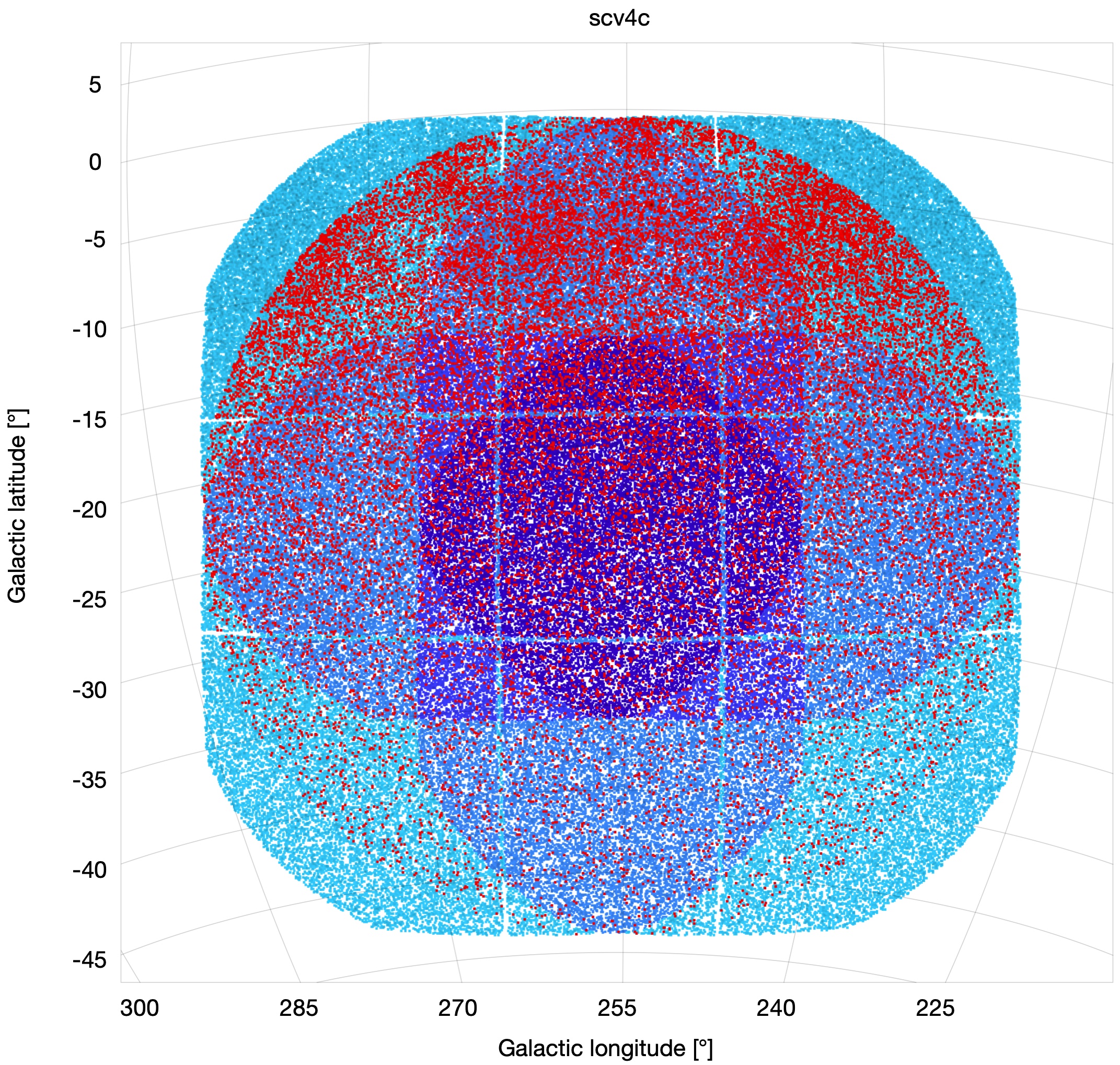}
\caption{Position of the {\tt scv4c} targets in LOPS2. Colour-code as in Figure \ref{fig:scvpic-sky}.}
\label{fig:LOPS_scv4c}
\end{figure}

\begin{figure}[h]
\centering
\includegraphics[width=0.49\textwidth]{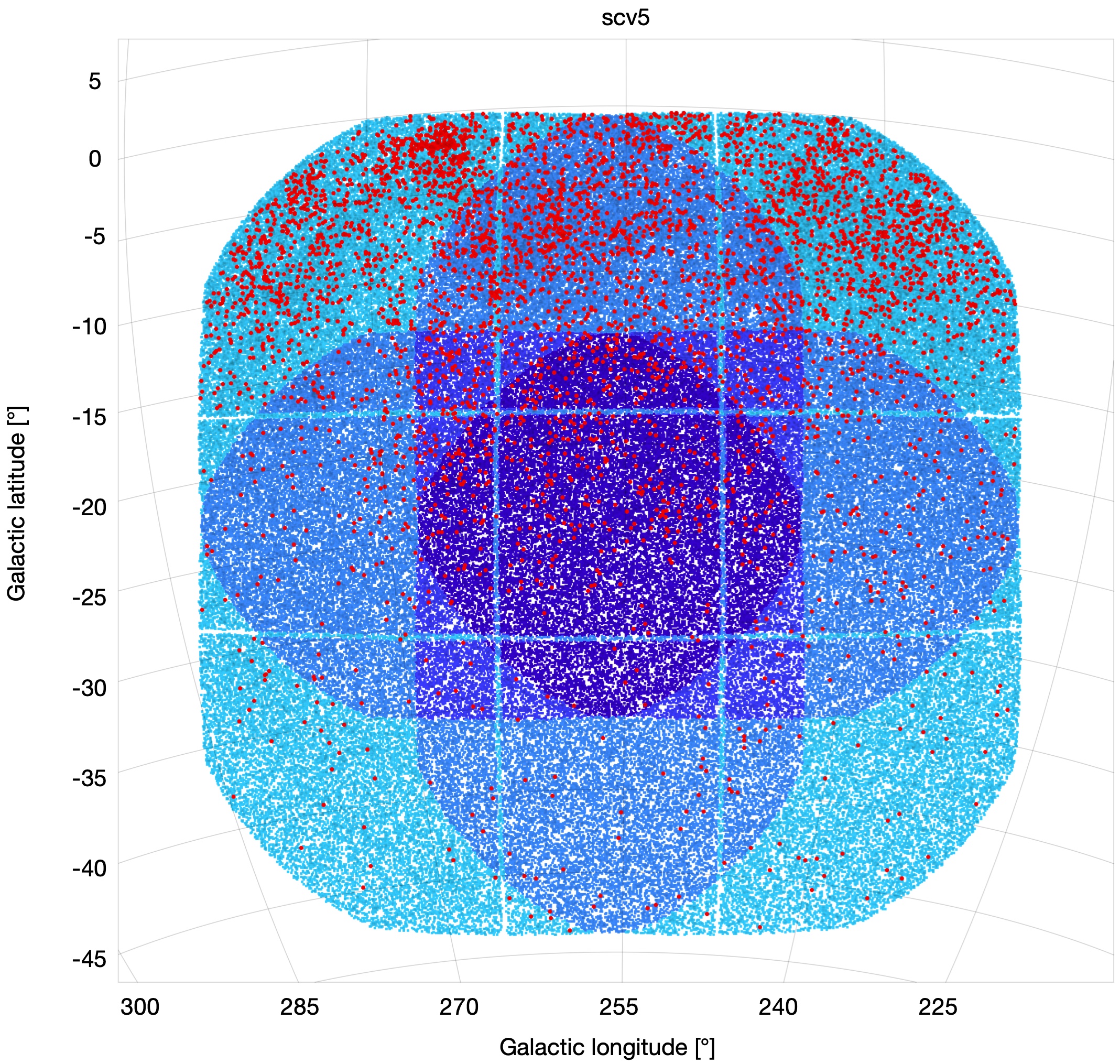}
\includegraphics[width=0.49\textwidth]{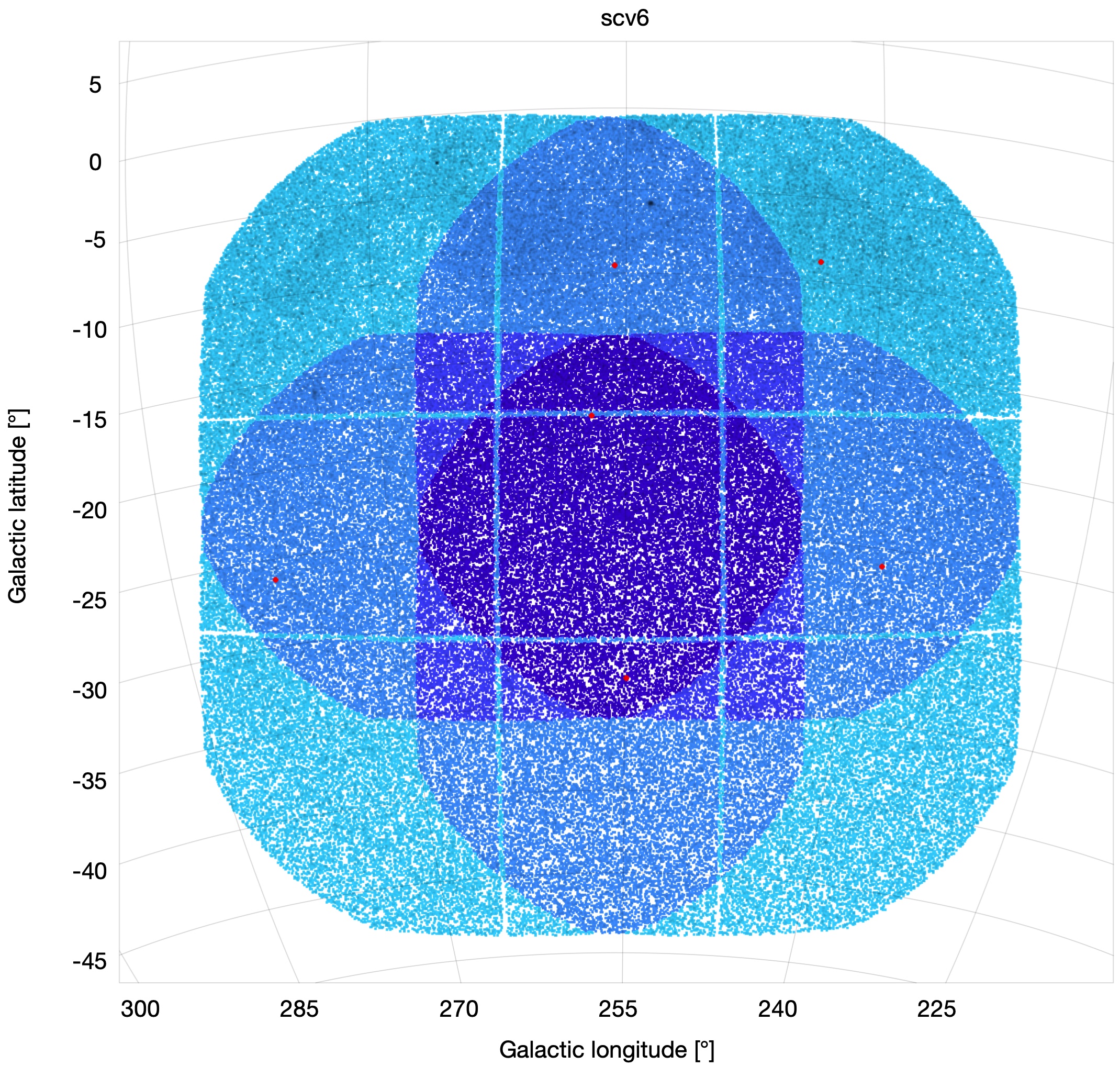}
\caption{Positions of the {\tt scv5} (left) and {\tt scv6} (right) targets in LOPS2. Colour-code as in Figure \ref{fig:scvpic-sky}.}
\label{fig:LOPS_scv56}
\end{figure}

\end{appendices}

\clearpage

\bibliography{scvPIC}

\end{document}